\documentclass[prx,amsmath,amssymb,aps,reprint,groupedaddress,raggedbottom,floatfix]{revtex4-2}

\usepackage{times}
\usepackage{graphicx}
\usepackage{amsmath}
\usepackage{amsfonts}
\usepackage{amssymb}
\usepackage{graphicx}
\usepackage{mathrsfs}
\usepackage{color}
\usepackage{xcolor}
\usepackage{hyperref}
\hypersetup{
	colorlinks,
	linkcolor={red!80!black},
	citecolor={black!10!blue},
	urlcolor={blue!80!black}
}

\usepackage{times}
\usepackage{graphicx}
\usepackage{amsmath}
\usepackage{amsfonts}
\usepackage{amssymb}
\usepackage{graphicx}
\usepackage[caption=false]{subfig}
\usepackage{mathrsfs}
\usepackage{color}
\newcommand{\p}{\partial}
\newcommand{\f}{\frac}
\newcommand{\md}{\mathrm{d}}

\newcommand{\Teff}{T_{\text{{eff}}}}

\begin{document}
\title{Dynamical heterogeneity in active glasses is inherently different from its equilibrium behavior}

	
\author{Kallol Paul}
\thanks{Contributed equally}
	\affiliation{TIFR Center for Interdisciplinary Science, Tata Institute of Fundamental Research, 36/P Gopanpally Village, Serilingampally Mandal, RR District, Hyderabad, 500046, Telangana, India}	
	\author{Anoop Mutneja}
	\thanks{Contributed equally}
	\affiliation{TIFR Center for Interdisciplinary Science, Tata Institute of Fundamental Research, 36/P Gopanpally Village, Serilingampally Mandal, RR District, Hyderabad, 500046, Telangana, India}
	\author{Saroj Kumar Nandi} 
	\affiliation{TIFR Center for Interdisciplinary Science, Tata Institute of Fundamental Research, 36/P Gopanpally Village, Serilingampally Mandal, RR District, Hyderabad, 500046, Telangana, India}
	\author{Smarajit Karmakar}
	\email{smarajit@tifrh.res.in}
	\affiliation{TIFR Center for Interdisciplinary Science, Tata Institute of Fundamental Research, 36/P Gopanpally Village, Serilingampally Mandal, RR District, Hyderabad, 500046, Telangana, India}






\begin{abstract}
Activity-driven glassy dynamics, while ubiquitous in collective cell migration, intracellular transport, dynamics in bacterial and ant colonies, etc, also extends the scope and extent of the as-yet mysterious physics of glass transition. 
Active glasses are hitherto assumed to be qualitatively similar to their equilibrium counterparts at an effective temperature, $\Teff$. Here we combine large-scale simulations and an analytical mode-coupling theory (MCT) for such systems and show that, in fact, an active glass is inherently different from an equilibrium glass. Although the relaxation dynamics can be equilibrium-like at a $\Teff$, effects of activity on the dynamical heterogeneity (DH), which has emerged as a cornerstone of glassy dynamics, are quite nontrivial and complex. With no preexisting data, we employ four distinct methods for reliable estimates of the DH length scales. Our work shows active glasses exhibit dramatic growth of DH and systems with similar relaxation times and $\Teff$ can have widely varying DH. To theoretically study DH, we extend active MCT and find excellent agreement between the theory and simulation results. 
Our results question the supposedly central role of DH in glassy dynamics and can have fundamental significance even in equilibrium.
\end{abstract}



\maketitle

Active glass characterizes the extreme dynamical slowdown without any discernible structural change or phase transition \cite{giulioreview,angelini2011} in a dense active system of self-propelled particles (SPP) with a self-propulsion force, $f_0$, and a persistence time, $\tau_p$, of their motion. It is an intriguing problem of statistical physics. On the one hand, several recent experiments show that signatures of glassy dynamics, such as non-exponential relaxation, caging, and dynamical heterogeneity, are crucial in many biological and biology-inspired systems, whose good minimalist  model is often a dense active matter of SPPs \cite{sriramreview,sriramrmp}. Examples include a cellular monolayer \cite{angelini2011,garcia2015}, cell-cytoplasm \cite{parry2014,sadati2014,zhou2009,nishizawa2017}, colonies of bacteria \cite{takatori2020} and ants \cite{gravish2013,gravish2015}, vertically vibrated rods \cite{dijksman2011,dauchot2005,deseigne2010}, and other artificial active systems \cite{dreyfus2005,palacci2010,klongvessa2019}. 
On the other hand, theoretical studies of such systems provide deeper insights into the physics of glasses as they extend the scope and extent of the problem via control parameters and emergent behaviors. Simulations have reproduced most experimental results \cite{ni2013,berthier2014,avila2014,mandal2016,flenner2016,mandal2017,mandal2020}, and analytical theories of equilibrium glasses, such as mode-coupling theory (MCT) \cite{berthier2013,activemct,liluashvili2017,feng2017,saroj2018,szamel2016} and random first-order transition theory (RFOT) \cite{activerfot}, have been extended for active glasses. 
Active systems from the perspective of single-particle dynamics \cite{chaki2020} and jamming transition \cite{ni2013,merrigan2020,bi2016} have also been studied. 
As a result, activity is known to drive the glass and jamming transitions to lower temperature or higher density. Notwithstanding some quantitative difference, such as the long-ranged velocity correlation \cite{szamel2016,caprini2020} or the evolving effective temperature, $\Teff$ \cite{activemct,berthier2013,cugliandolo2019}, active glasses are considered qualitatively similar to equilibrium glasses.
{However, this apparent similarity is somewhat puzzling as activity leads to nontrivial behaviors, such as flocking \cite{vicsek1995} and giant number fluctuation \cite{vijay2007,sriram2003,sriramrmp,sriramreview} in a dilute system, and calls for a rigorous theoretical exploration of the problem. }

In this work, we combine large scale molecular dynamics simulations with an analytical study via extended active-MCT \cite{activemct} formalism and show that, in fact, active glasses are qualitatively different from their equilibrium counterparts: whereas the relaxation dynamics is similar to that of equilibrium glasses, nontrivial effects of activity are observed on the dynamical heterogeneity (DH).
DH refers to the coexistence of heterogeneous, transient local dynamic behavior and has emerged as a salient feature of glassy dynamics \cite{ediger2000,berthier2011,giulioreview}. The existence of DH in glassy systems have been confirmed both in simulations \cite{yamamoto1998,franz2000,smarajitPNAS2009} and experiments \cite{weeks2000,dauchot2005}, and provides the much sought after growing length scale in the problem.
{It is important to emphasize here that there are two commonly talked about length scales in the literature, namely the DH length scale, which will be discussed here, and the point-to-set (PTS) length scale. PTS length scale is commonly thought of as the length scale that controls the relaxation time. In contrast, DH length scale has been shown to be directly related to the observed breakdown in Stokes-Einstein Relation, which relates the diffusion constant to the relaxation time in a simple liquid. The DH length scale is also believed to be associated with the well-known non-Gaussian dynamical behaviour observed at characteristic relaxation time scales. Recently, it is found to be linked to the fragility in the system; strong liquids which obey nearly Arrhenius-like temperature dependence of the relaxation time (viscosity) show much weaker growth of DH length, whereas fragile liquids which have strong non-Arrhenius temperature dependence of the relaxation time show much stronger growth of DH length scale. This also points to a close link between the PTS and DH length scale, which is certainly an interesting question to be addressed. Here we show that active glasses with the same relaxation time can have varying DH and show enhanced dynamic heterogeneity in the strong liquid limit at large activity in complete contrast with the equilibrium behaviour, suggesting a possible decoupling from the relaxation dynamics; this raises an interesting fundamental question on the presumed central role of DH in glassy dynamics.
}

\begin{figure*}[!htpb]
	\includegraphics[width=0.96\textwidth]{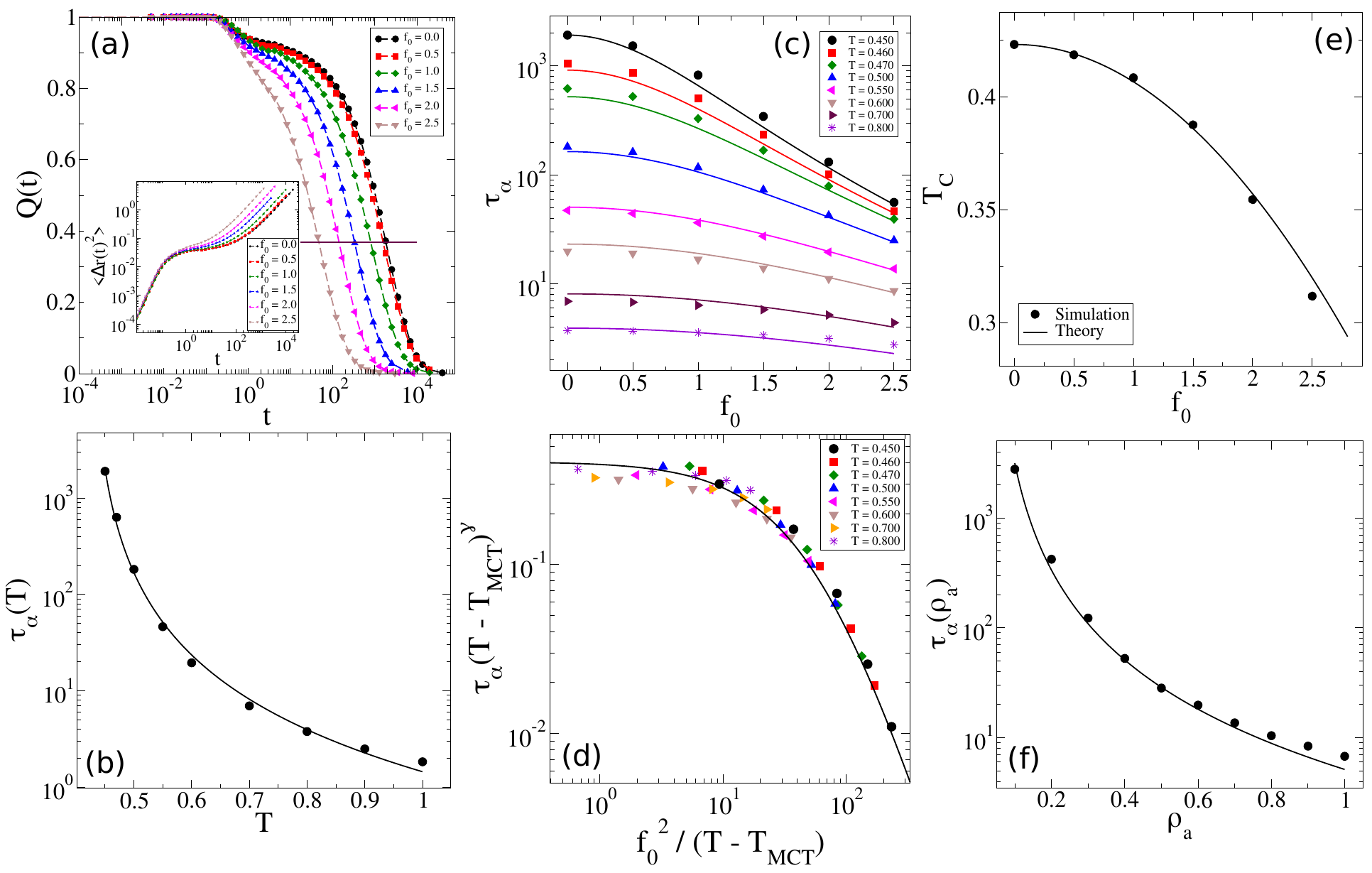}
	\caption{{\bf Characterization of relaxation dynamics}. (a) Decay of the overlap function, $Q(t)$, and growth of the MSD, $\langle \Delta r(t)^2\rangle$, (inset) become faster with increasing activity. (b) Symbols are data of $\tau_\alpha$ as a function of $T$ for a passive system, and the line is a fit with the MCT prediction $\tau_\alpha\sim (T-T_{MCT})^{-\gamma}$, that gives $T_{MCT}=0.423$ and $\gamma=2.35$. (c) $\tau_\alpha$ as a function of $f_0$ at different $T$, symbols are simulation data, and the lines are active MCT prediction (not individual fits). (d) Equation (\ref{tauscaling}) predicts a master curve when $\tau_\alpha(T-T_{MCT})^\gamma$ is plotted as a function of $f_0^2/(T-T_{MCT})$ at different $T$ and $f_0$: the simulation data (symbols) agree well with the theoretical prediction (line). {\bf Inset:} Comparison of $T_C$ between simulation and theory. (e) The MCT transition temperature, $T_C$, for an active system, the line is the theoretical prediction. (f) $\tau_\alpha$ for an active system at $T=0.41$ and $f_0=1.5$ with varying $\rho_a$. Symbols are simulation data and the line represent the theory, Eq. (\ref{rhoascaling}).}
	\label{taucomp}
\end{figure*}

Most works to date have focused on the relaxation dynamics in an active glass and only a handful of studies have \cite{angelini2011,garcia2015,malinverno2017,cerbino2021,flenner2016} investigated different aspects of DH revealing nontrivial effects of activity: experiments on active cellular monolayers \cite{malinverno2017,cerbino2021} show higher DH with increasing activity though the relaxation time becomes smaller, a trend opposite to what one expects for equilibrium glass-forming liquids. This exciting result, though puzzling, can be rationalized in the light of our work. We first show that the relaxation dynamics in an active glass is similar to that in an equilibrium system at a $\Teff$ (Fig. \ref{taucomp}), yet, the behavior of DH is drastically different (Fig. \ref{chi4_temp}). We then present reliable computations of length scales associated with DH in active systems via four distinct methods (Fig. \ref{lengthscale}) and, finally, illustrate the nontrivial aspects of the activity in a glassy system (Fig. \ref{sametau_activity}).

To test the generality of our results, we have simulated two distinct, well-known, three-dimensional models of glassy systems: the Kob-Andersen binary mixture (3dKA) with $80:20$ particle ratio and another binary mixture (3dHP) with $50:50$ particle ratio. We present the results of the 3dKA model in the main text and some results of the 3dHP model in SM, Fig. S11. To include activity in our models, we randomly select $\rho_a=10\%$ of the particles and specify them as active. The active particles are then applied constant forces of magnitude $f_0$ in random directions keeping the vector sum to zero; this ensures momentum conservation of the centre of mass (CoM) in our simulations. We reshuffle the directions of active forces after every $\tau_p$ time. Our choices of $\rho_a$ and the model of activity are motivated by intracellular systems, such as the cytoplasm or cellular cortex \cite{jacques2015}, where only a fraction of particles are active. Unless otherwise stated, we use $\rho_a=0.1$, though we also show that our results are generic for other values of $\rho_a$.
{We keep density and $\tau_p=1.0$ constant for most of our presented data and study the glassy properties as functions of temperature, $T$, and $f_0$. We show 
some results of effect of $\tau_p$ variation on DH, although variation of $\tau_p$ is kept small as larger values of $\tau_p$ can altogether alter the glassy dynamics as reported in \cite{mandal2020}.} We also extend active MCT, presented in Ref. \cite{activemct}, to investigate DH following the IMCT formalism presented in \cite{IMCT}. Further details of the models and the simulations are given in the materials and methods and the SM. We show the detailed theoretical calculations in the SM.

\section*{Results}

\subsection*{The relaxation dynamics}
We first show that the relaxation dynamics of an active glass is similar to that of an equilibrium system at a suitably chosen $\Teff$. The overlap function, $Q(t)$, and the mean-squared displacement (MSD), $\langle \Delta r(t)^2\rangle$, are shown in Fig. \ref{taucomp}(a) and in the inset respectively, as a function of time $t$ (see Materials and Methods for definitions). At a particular $T=0.45$, both $Q(t)$ and MSD for the passive system ($f_0=0$) show the plateau, characteristic of glassy dynamics. As activity increases from $0$, decay of $Q(t)$ and growth of MSD become faster, consistent with earlier simulations \cite{berthier2014,flenner2016,mandal2016} and experiments \cite{klongvessa2019} (note that we are varying activity via $f_0$ alone keeping $\tau_p=1.0$ fixed). The 
relaxation time, $\tau_\alpha$, is defined as $Q(\tau_\alpha)=1/e$. Active MCT \cite{activemct} predicts 
\begin{equation}\label{tauscaling}
\tau_\alpha\sim |T-T_{MCT}+Kf_0^2|^{-\gamma}
\end{equation}
where $\gamma$ is the same exponent as in a passive system and $T_{MCT}$ is the MCT transition temperature, where $\tau_\alpha$ diverges within the MCT formalism in the absence of activity. $K$ is a constant. For the particular model of activity in this work, defined as SNTC in Ref. \cite{activemct}, we have $K=H\tau_p/(1+G\tau_p)$, with $H$ and $G$ being two constants. We show the data of $\tau_\alpha(T)$ for the equilibrium 3dKA model in Fig. \ref{taucomp} (b) (all simulation data presented here are for this model, results for the 3dHP model are shown in SM Fig. S11). A fit of the MCT prediction, $\tau_\alpha\sim |T-T_{MCT}|^{-\gamma}$, with the simulation data gives $\gamma\simeq2.35$ and $T_{MCT}\simeq 0.423$, consistent with existing studies \cite{kob1995,ashwin2003}. Using these values, we fit one set of data for the active system and obtain $K\simeq 0.0155$. 
Note that once $K$ is determined using one set of data, there are no other free parameters within the theory.
We then plot Eq. \ref{tauscaling} along with the simulation data in Fig. \ref{taucomp}(c) and find remarkable agreement. 
Moreover, Eq. \ref{tauscaling} shows that plot of $\mathcal{Y}=\tau_\alpha(T-T_{MCT})^\gamma$  as a function of $\mathcal{X}=f_0^2/(T-T_{MCT})$ at different $f_0$ and $T$ should follow a master curve: $\mathcal{Y}\sim (1+K\mathcal{X})^{-\gamma}$. Figure~\ref{taucomp}(d) shows that simulation data agree with this MCT result. 
Equation (\ref{tauscaling}) also predicts the MCT transition temperature, $T_C$, for the active system $T_{C}=T_{MCT}-Kf_0^2$ (Fig. \ref{taucomp}e). 

We have also carried out simulations at different values of $\rho_a$ and find that the qualitative behavior remains the same. Since $K$ represents the effect of activity, we expect it to be proportional to $\rho_a$, thus $K=\tilde{K}\rho_a$, where $\tilde{K}$ is a constant. Since $K=0.0155$ when $\rho_a=0.1$, we obtain $\tilde{K}=0.155$. Thus, Eq. (\ref{tauscaling}) becomes
\begin{equation}\label{rhoascaling}
\tau_\alpha\sim |T-T_{MCT}+\rho_a \tilde{K}f_0^2|^{-\gamma},
\end{equation}
where all the parameters are already determined. Figure \ref{taucomp}(f) shows simulation results for $\tau_\alpha$ at $T=0.41$ and $f_0=1.5$ as a function of $\rho_a$, and the line is the theoretical prediction, Eq. (\ref{rhoascaling}). Thus, $\rho_a$ only scales the activity parameters without affecting the qualitative behavior.
The excellent agreement between the theory and simulation data with $\gamma$ being the same exponent as for the passive system shows that the relaxation dynamics in an active glass is similar to that in an equilibrium glass at  $\Teff=T+Kf_0^2$. Despite this, we now show that active glasses are qualitatively different from equilibrium glasses: the nontrivial effects of activity manifest in DH characterized via four-point correlation function, $\chi_4(t)$. We first present the extension of active MCT to study DH and then the simulation results for DH.

\subsection*{Active In-homogeneous Mode-Coupling Theory (Active-IMCT) }
To the best of our knowledge, MCT for an active system has not been extended for the DH. We present the detailed derivation in the SM Sec. S8 and outline the main result here. We follow the IMCT formalism, developed in Ref. \cite{IMCT}, and obtain the four-point correlation function in terms of a corresponding susceptibility via linear response theory. 
Within our theory, activity enters as a colored noise in the continuity equation of momentum density \cite{activemct}.
Since an active system is inherently out of equilibrium, one must study both the correlation and (integrated) response functions, $Q(t)$ and $F(t)$, respectively, as well as their corresponding susceptibilities. As detailed in the SM, the equations of motion for the susceptibilities, $\chi_Q(t)$ and $\chi_F(t)$ are 
\begin{subequations}
\label{chieq}
\begin{align}
\f{\p\chi_Q(t)}{\p t} &=\nu(t)+(1+\zeta)Q(t)-(T-p)\chi_Q(t) \nonumber\\
-&\int_0^tm(t-s)\f{\p \chi_Q(t)}{\p s}\md s-\int_0^t \Sigma(t-s)\f{\p Q(s)}{\p s}\md s \\
\f{\p\chi_F(t)}{\p t} &=(1+\zeta)F(t)-(T-p)\chi_F(t) \nonumber\\
-&\int_0^tm(t-s)\f{\p \chi_F(t)}{\p s}\md s-\int_0^t \Sigma(t-s)\f{\p F(s)}{\p s}\md s,
\end{align}
\end{subequations}
\begin{align}
\text{with, }\nu(t)&=-\int_t^\infty \Delta(s) \f{\p\chi_F(s-t)}{\p s}\md s, \\
\zeta=\int_0^\infty &\Delta(s) \f{\p\chi_F(s)}{\p s}\md s, \,\,\, p=\int_0^\infty \Delta(s)\f{\p F(s)}{\p s}\md s\\
m(t)&=2\lambda \frac{Q(t)^2}{T_{{eff}}(t)},\\
\Sigma(t)&=2\lambda\f{Q(t)\chi_Q(t)}{T_{{eff}}(t)}+2\lambda\f{Q(t)^2}{T_{{eff}}(t)^2}\kappa f_0^2, \label{sigmaterm}\\
\text{and} \,\,\,\Teff(t)&=\frac{\p Q(t)}{\p t}\bigg/\frac{\p F(t)}{\p t},
\end{align}
where $\Delta(s)$ is the active force statistics: $\Delta(s)=f_0^2\exp[-s/\tau_p]$. $\kappa$ is a system-dependent constant ({\color{blue}see SM}), set to $\kappa=2.0$ in this work; a more detailed analysis will be presented elsewhere. As shown in Ref. \cite{IMCT}, $\chi_Q(t)\equiv \chi_4(t)$ is the desired four-point correlation function. These equations must be simultaneously solved along with the equations for $Q(t)$ and $F(t)$, as presented in Ref. \cite{activemct} and the SM.

\subsection*{Dynamical heterogeneity in active glasses}
\begin{figure}[!h]
 \includegraphics[width = 0.47\textwidth]{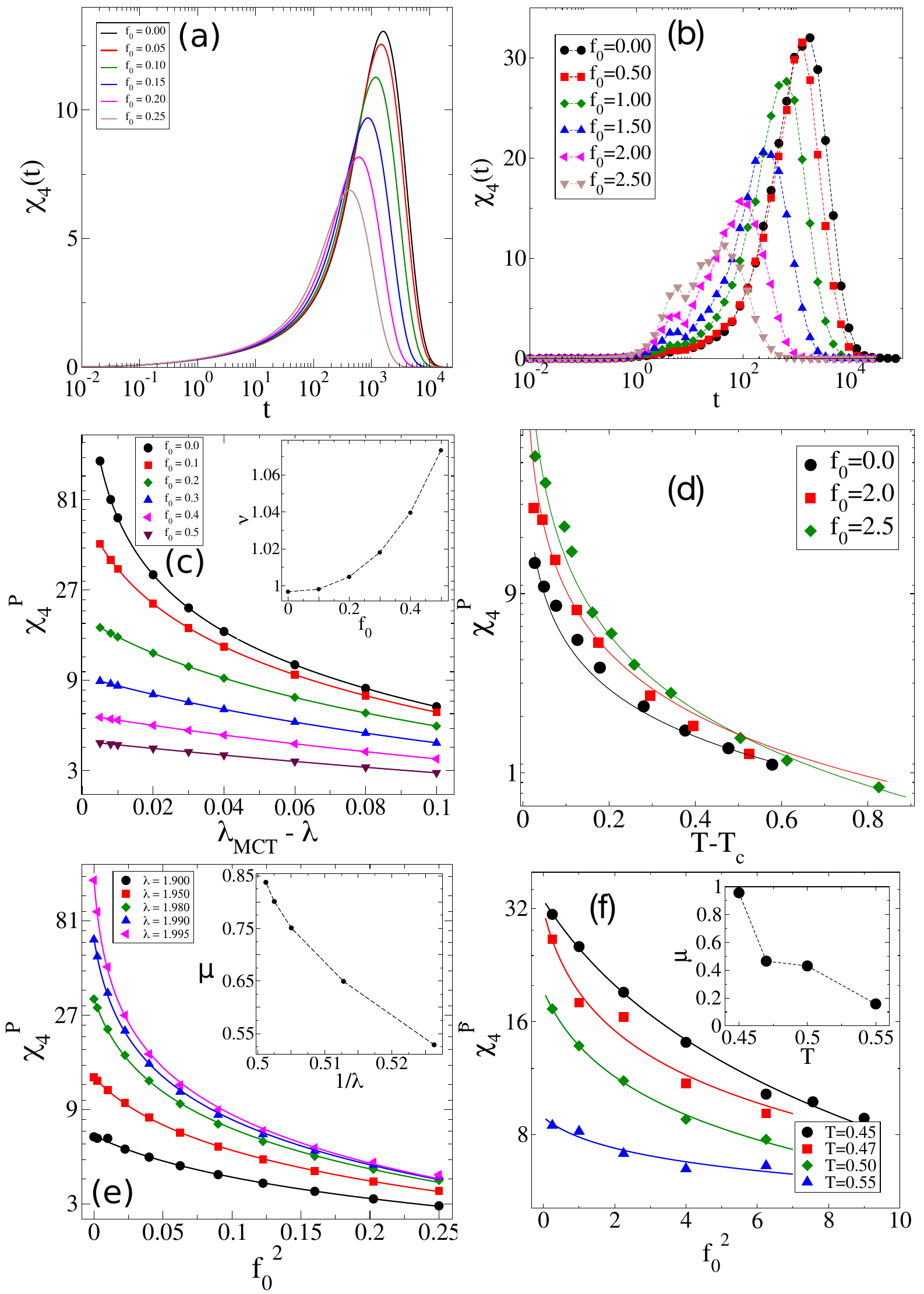}
 \caption{{\bf Nontrivial effects of activity on dynamical heterogeneity}. (a) Active-IMCT results for $\chi_4(t)$ at different $f_0$ and $\lambda=1.95$. $\chi_4^P$, the peak value of $\chi_4(t)$, decreases with increasing activity. (b) Simulation data for $\chi_4(t)$ at different $f_0$ and $T=0.45$ also shows similar behavior as in active-IMCT.
 (c) $\chi_4^P$ as a function of $\lambda_{MCT}-\lambda$ at different $f_0$. Symbols are numerical solutions of active-IMCT, and the lines are fits with a power-law (see text). Surprisingly, the exponent of the power-law behavior increases with $f_0$. (d) Simulation results for $\chi_4^P$ as a function of $T$ at different $f_0$, and lines represent fits with a power-law where the exponent depends on $f_0$ (see text). (e) 
 Symbols are active-IMCT solutions for $\chi_4^P$ as a function of $f_0^2$ at different $\lambda$, fits with power-law behavior shows the exponent depends on $\lambda$. (f) Similar behavior as in (e) is also observed in simulations; fitting the simulation data (symbols) with a power-law form gives the exponent $\mu$ that depends on $T$.}
 \label{chi4_temp}
\end{figure}
We now discuss the effects of activity in DH, characterized via $\chi_4(t)$, defined in the Materials and Methods. Figure \ref{chi4_temp}(a) shows the numerical solutions from IMCT, and Fig. \ref{chi4_temp}(b) shows the simulation results for  $\chi_4(t)$ at different $f_0$. Both within IMCT and simulations, $\chi_4^P$, the peak value of $\chi_4(t)$, decreases with increasing $f_0$. 
Figure~\ref{taucomp}(c) shows that increase in $f_0$ results in decreasing the $\tau_\alpha$ and driving the system away from the glassy regime; a decrease of $\chi_4^P$ with an increase in $f_0$ is, therefore, expected. But the trend is different from equilibrium-like behavior. Figures \ref{chi4_temp}(c) shows the IMCT results for $\chi_4^P$ at different $f_0$ as a function of $\lambda_{MCT}-\lambda$, where $\lambda_{MCT}=2.0$ is the equilibrium critical point. Since MCT is a critical phenomena like theory, we expect a power-law behavior: $\chi_4^P\sim (\lambda_C-\lambda)^{-\nu}$, where $\lambda_C$ is the critical point in the presence of activity. A fit of the data with this form shows a small $f_0$-dependence of $\nu$. 
Figure \ref{chi4_temp}(d) show the simulation results of $\chi_4^P$ as a function of $T$ at three $f_0$ including $f_0=0$. {The lines are fits with a function $\chi_4^P\sim (T-T_C)^{-\nu}$. Consistent with IMCT, we find the exponent $\nu$ depends on activity: $0.84$ for the passive system, whereas $\nu=1.10$ and $1.33$ 
for $f_0=2.0$ and $2.5$ respectively. In contrast to our findings, one expects $\nu$ to remain constant if the active glass resembles an equilibrium system at a suitably defined $\Teff$.} Note one obtains three-point susceptibility from IMCT theory but it is well-known that three-point and four-point have the same scaling behaviour near the MCT transition temperature. In this study, we computed the three-point susceptibility as a proxy for four-point susceptibility to compare the theory with the simulation results.

To further probe this effect of activity of DH, we now study $\chi_4^P$ as a function of $f_0$ at different $\lambda$ or $T$ within IMCT (Fig. \ref{chi4_temp}e) and in simulations (Fig. \ref{chi4_temp}f), respectively. A fit of the function $\chi_4^P\sim (1+bf_0^2)^{-\mu}$, where $b$ is a constant, with the simulation data shows $\mu$ depends on $\lambda$ or $T$ [insets, Figs. \ref{chi4_temp}(e) and \ref{chi4_temp}(f)]; $\mu$ decreases as the passive system goes away from the critical point. Thus, varying activity not only changes $T_C$ but also affects the exponents of the power law.
The time $\tau_{peak}$, when $\chi_4(t)$ attains its maximum, provides a measure of a relaxation time. As shown in the SM Fig. S9, $\tau_{peak}$ is proportional to $\tau_\alpha$, demonstrating that the relaxation dynamics, characterized via either of $Q(t)$ or $\chi_4(t)$, can be described in terms of a $\Teff$ (Figs. \ref{taucomp}c and \ref{taucomp}d). Thus, activity has distinctive effects on relaxation dynamics and DH; it can be treated in terms of a $\Teff$ for the former but not for the latter. Besides, within our extended MCT, the nontrivial effect of activity on the DH comes from the last term in Eq. (\ref{sigmaterm}). This term is a direct consequence of the nonequilibrium nature of the system and is absent in equilibrium. In any case, the distinctive effects of activity on the relaxation dynamics and DH length scale implies a possible decoupling from each other.


\subsection*{Dynamical Heterogeneity Length Scale}
Having discussed the non-trivial effects of activity on the DH, we now compute the dynamic length scale, $\xi_D$, associated with DH itself. Reliable estimates of $\xi_D$ in model active glassy systems are rare \cite{ghoshal2020}; in this work, we compute and compare $\xi_D$ via four different procedures - (i) finite size block analysis scaling of $\chi_4$, (ii) block analysis of van Hove function, (iii) spatial correlation of displacements at structural relaxation time, and (iv) via the wave vector ($q$) dependence of four-point structure factor, $S_4(q,t)$.
The first two procedures use the method of block analysis following Refs. \cite{saurish2017,bhanu2018}.
The simplicity and computational efficiency of the method make it immediately useful for our current study as we must compute $\xi_D$ not only as a function of $T$ but also as a function of $f_0$. The excellent agreement of the results from the four methods shows the reliability of our results.

 \begin{figure}
 	\begin{center}
 		\includegraphics[width=0.5\textwidth]{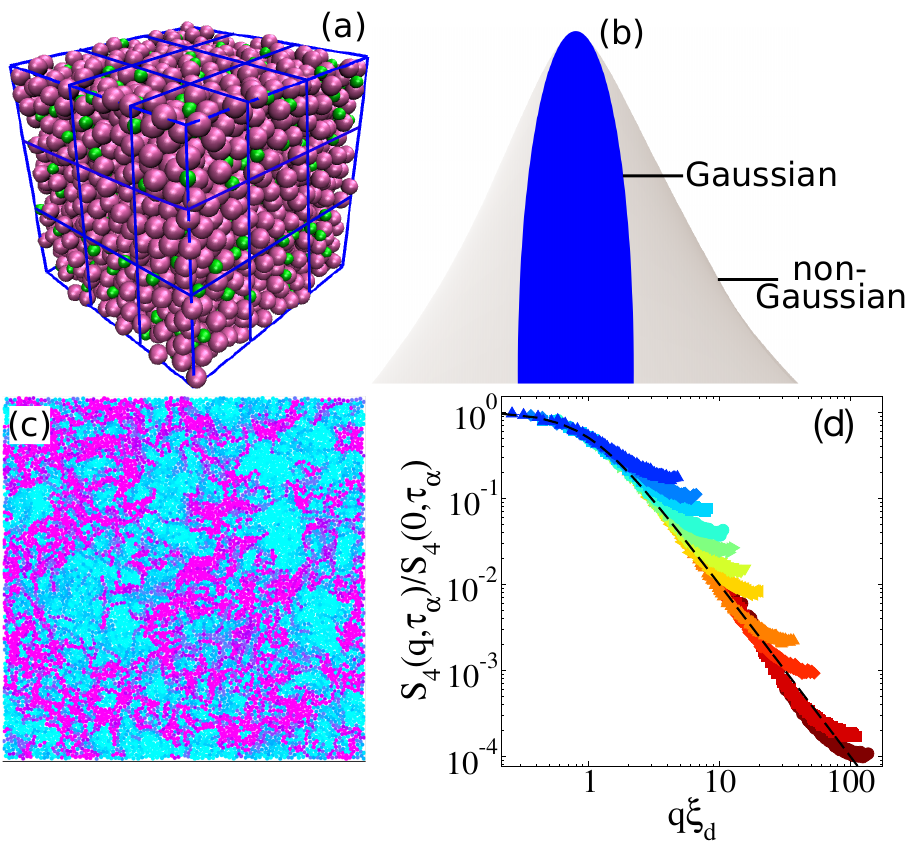}
 		\caption{Schematic illustration of the four methods to reliably compute the dynamic heterogeneity length scale. (a) Finite size block analysis scaling of $\chi_4$. (b) Block analysis of the non-Gaussian nature of van Hove function. (c) Spatial correlations of displacements at a time scale of relation time. (d) Scaling analysis of the four-point structure factor, $S_4(q,t)$.}
 		\label{schematicmethod}
 	\end{center}
 \end{figure}
 
 {
In Fig. \ref{schematicmethod}, we only schematically illustrate the four methods and refer the reader to Appendix \ref{comp_xi} for a detailed description and the SM for further technical details. In the block analysis method, a large system with linear size $L$ is simulated, then divided into smaller blocks, as schematically shown in Fig. \ref{schematicmethod}(a). Then one computes both $Q(t)$ and $\chi_4(t)$ as a function of the linear size of the blocks. The primary advantages of computing both these correlation functions are the better averaging over a large number of blocks and inclusions of all possible fluctuations that contribute to the computation of $\chi_4(t)$, {\it e.g.} density, composition, temperature and activity fluctuations \cite{flenner2010,saurish2017} as each of the blocks can be thought of as being submerged in a bigger bath similar to grand canonical ensembles. This is a modified version of the usual finite-size scaling methods to apply to glassy systems \cite{saurish2017,flenner2010}. In previous studies \cite{saurish2017,bhanu2018} in equilibrium glassy systems, the method's usefulness is demonstrated over a wide range of model systems across dimensions.

\begin{figure*}
	\begin{center}
		\includegraphics[width=0.97\textwidth]{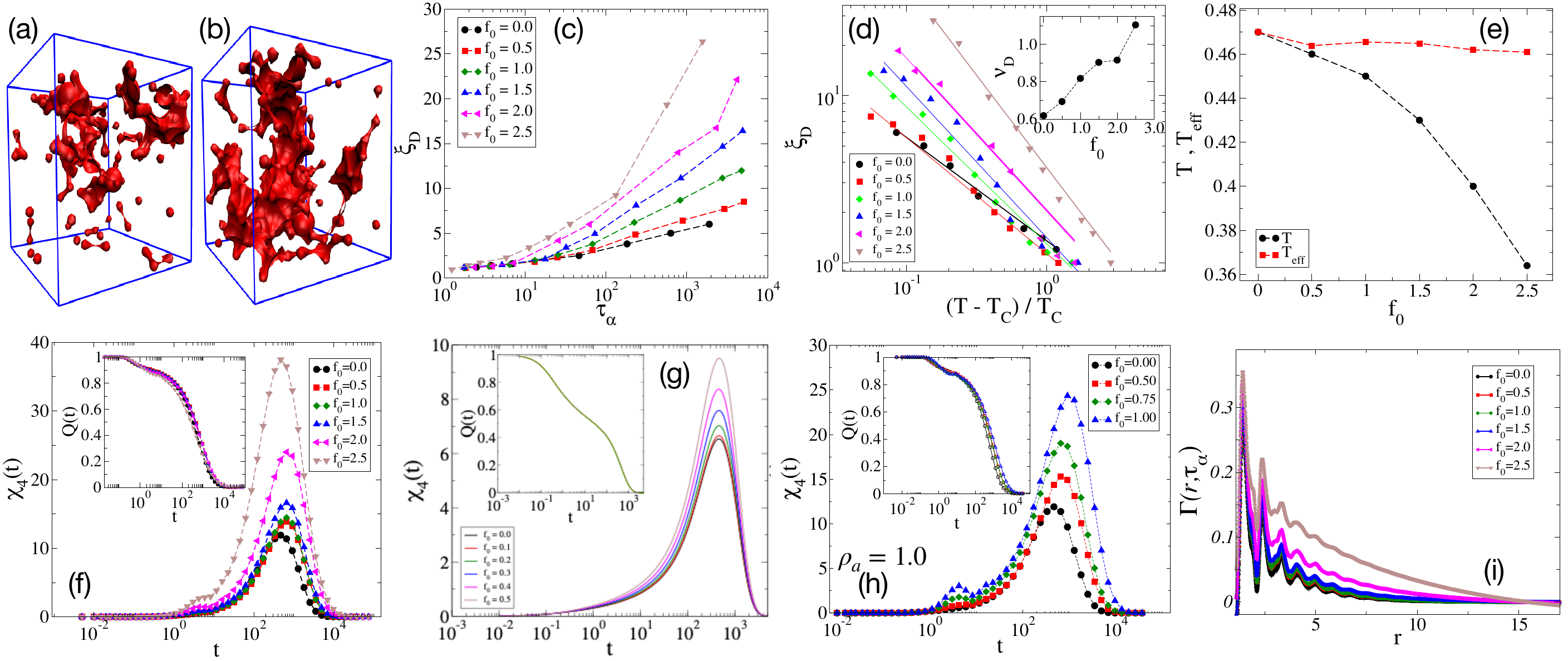}
		\caption{{\bf Dramatic effects of activity on dynamical heterogeneity}. (a-b) Qualitative measures of CRRs showing the fraction of fast relaxing particles in a passive (a) and an active system (b), at the same $\tau_\alpha=10^3$. CRR provides a typical estimate of DH, and the comparison shows larger DH in an active system. (c) $\xi_D$ as a function of $\tau_\alpha$ at different $f_0$. The data shows that for a given $\tau_\alpha$, a system with higher $f_0$ has a larger $\xi_D$. (d) The fit of simulation data (symbols) for $\xi_D$ with the power-law prediction of MCT (lines); the exponent $\gamma_D$ is expected to be constant for effective equilibrium-like behavior. However, we find that $\gamma_D$ almost linearly increases with $f_0$ (inset). (e) To obtain comparable $\tau_\alpha$, we choose a set of $T$ and $f_0$ such that $\Teff$ remains similar. (f) $\chi_4(t)$ for the parameters as in (e); a system with larger $f_0$ has higher $\chi_4^P$. {\bf Inset:} All these systems have similar $\tau_\alpha$ as seen from the plot of $Q(t)$. (g) $\chi_4(t)$, obtained from active-IMCT, also has a similar behavior as in simulation, i.e., higher $\chi_4^P$ for larger $f_0$ when $Q(t)$, and hence $\tau_\alpha$, are similar (inset). The parameter $\lambda=1.9$ when $f_0=0$ and $\lambda$ for other values of $f_0$ are chosen to overlap $Q(t)$ with that of the passive system. (h) Similar results as in (f) but with $\rho_a=1.0$. (i) Increasing spatial correlation,
			as measured by the excess displacement-displacement correlation function, $\Gamma(r,\tau_\alpha)$, with increasing activity. These correlation functions are again computed at the same $\tau_\alpha$ to highlight the growth of spatial heterogeneity with increasing activity (see text for details).}  
		\label{sametau_activity}
	\end{center}
\end{figure*}

The second method (Fig. \ref{schematicmethod}b) relies on the non-Gaussian behavior of the self part of the van Hove function, $G_s(x,\tau_\alpha)$ (defined in Method section and elaborated in SM) in a glassy system. The method for an equilibrium system was demonstrated in Ref. \cite{bhanu2018}. It is well-known that the van Hove function shows universal non-Gaussian beahviour in supercooled regimes, and it is a hallmark signature of underlying dynamic heterogeneity in the system. The basic idea is to study the van Hove function as a function of spatial coarse-graining scale. In this method, one recomputes the displacement of a coarse-grained region of size $L_B$ and then computes the van Hove function from the coarse-grained displacement field. As the coarse-graining length scale crosses the typical dynamic length $\xi_D$, the nature of $G_s(x,\tau_\alpha)$ should change from non-Gaussian to Gaussian as the system becomes eventually spatially homogeneous above this length scale. Thus by computing the Kurtosis of the van Hove function with increasing coarse-graining length, $L_B$, one can estimate the typical dynamic length scale, $\xi_D$, via scaling analysis as elaborated in the Method section.

The third method relies on the study of spatial correlation of the displacement fields of the particles over the structural relaxation time scale \cite{indrajit2020} as shown in Fig. \ref{schematicmethod}c. One can see that if one computes the spatial displacement-displacement correlation between two particles separated by a distance $r$, $\Gamma(r,\tau_\alpha)$, (see Method section for definition) then one will be able to estimate the heterogeneity length scale by checking how fast the correlation function decay as a function of spatial distance. This method has been shown to be very robust in estimating the dynamic heterogeneity length scale, $\xi_D$, in equilibrium supercooled liquids as elaborated in \cite{indrajit2020} and reference therein. In this work, we show that the same method works equally well for non-equilibrium conditions as it gives an estimate of the correlation length, which is very similar to the one obtained using methods discussed in the previous section.

Finally, the fourth method relies on the spatial correlation of the displacements obtained via a four-point structure factor. This is a standard method for equilibrium systems \cite{flenner2010}. In this method, one computes the four-point dynamic susceptibility, $S_4(q,\tau_\alpha)$ at typical relaxation time where one expects the heterogeneity to be maximum and then uses Ornstein-Zernike (OZ) liquid state theory to collapse all the data and obtain the correlation length $\xi_D$ as shown in Fig. \ref{schematicmethod}d. The OZ theory suggests that the inverse of $S_4(q,\tau_\alpha)$ will have quadratic wave vector, $q$, dependence. The line passing through the data points is the prediction of the OZ theory. Thus an appropriate scaling analysis can be used to obtain the length scale as elaborated in the Method section. We want to highlight that this method can suffer from finite size effects, and one needs to work with large system sizes compared to the typical length scale to get reliable estimates of the length scale ~\cite{PhysRevLett.105.015701}. As shown in the Appendix \ref{comp_xi}, $\xi_D$ computed via these distinct methods compare well with each other; therefore, we are confident of the length scales reported in this work.
}

\subsection*{Dramatic Effects of Activity on Dynamical Heterogeneity}
Within the DH picture of equilibrium glassy dynamics, $\xi_D$ is believed to control the relaxation dynamics  \cite{giulioreview,ediger2000}. We have already shown that the nature of DH in an active glass is different from equilibrium-like behavior (Fig.~\ref{chi4_temp}). This nontrivial behavior, within active-IMCT, is governed by the second term in the memory kernel, Eq. (\ref{sigmaterm}).
We now demonstrate the striking effects of activity by considering different systems whose parameters are such that the typical relaxation times remain the same.
As DH monotonically relates to $\tau_\alpha$ in an equilibrium system \cite{giulioreview,berthier2011}, the former should be the same when the latter remains unchanged.

However, the scenario for an active glass, surprisingly, turns out to be drastically different. 
DH essentially refers to the coexistence of dynamic fast and slow-moving regions in the same system. A set of adjacent particles with similar $\tau_\alpha$ relaxes collectively and is known as cooperatively rearranging region (CRR) that gives a measure of DH \cite{patrickroyall2017}. Figures~\ref{sametau_activity}(a) and \ref{sametau_activity}(b) show typical estimates of CRRs of the fast-moving particles in an equilibrium system and an active system, respectively, both having the same $\tau_\alpha=10^3$ ({\color{blue}see SM, Sec. S6 for details}). At this particular $\tau_\alpha$, the CRR for the equilibrium system is disconnected whereas that for the active system is system-spanning, showing higher DH in an active system at the same $\tau_\alpha$. For more quantitative analysis, we present $\xi_D$ as a function of $\tau_\alpha$ in Fig. \ref{sametau_activity}(c); at a given $\tau_\alpha$, a system with higher activity (that is larger $f_0$) has larger $\xi_D$. 
Note the drastic growth of $\xi_D$ with activity: within a similar window of $\tau_\alpha$, compared to its high-$T$ value, $\xi_D$ in the highest active system grows by a factor of $\sim 30$ whereas that in the passive system grows by a factor of merely $4-5$.

The nontrivial nature of activity also manifests via the activity-dependence of the exponents in the growth-laws of $\xi_D$.
MCT predicts power-law divergence of $\xi_D$ at $T_C$, obtained from fitting the data of $\tau_\alpha$ (Fig. \ref{taucomp}e). Figure~\ref{sametau_activity}(d) shows simulation data of $\xi_D$ as a function of $(T-T_C)/T_{C}$, where the lines represent fits with the function $\xi_D\sim [(T-T_C)/T_C]^{-\nu_D}$; the data agree well with the power-law form in the range of $T$ and $f_0$ in the regime of the simulations. An effective-equilibrium-like description implies the same $\nu_D$ at different activity, as in the relaxation dynamics, Eq. (\ref{tauscaling}).
However, consistent with the behavior of $\chi_4^P$ (Fig. \ref{chi4_temp}), we find $\nu_D$ depends on activity and almost linearly increases with $f_0$, as shown in the inset of Fig. \ref{sametau_activity}(d).

We now show that different active systems with the same $\tau_\alpha$ can have distinct DH suggesting a decoupling of the two in the presence of activity. Figure \ref{sametau_activity}(f) shows simulation data for $Q(t)$ and $\chi_4(t)$ at different $T$ and $f_0$. The parameters are such that all the systems have very similar $\tau_\alpha$,  as is evident from the plots of $Q(t)$ shown in the inset of Fig. \ref{sametau_activity}(f). Analysis of $\tau_\alpha$ in terms of $\Teff$ implies all the systems have similar $\Teff$, as shown in Fig. \ref{sametau_activity}(e) that also shows the values of $T$ at different $f_0$. 
A monotonic relation between the DH and the relaxation dynamics, akin to an equilibrium system, implies overlapping $\chi_4(t)$ for all the systems. However, Fig. \ref{sametau_activity}(f) shows that $\chi_4(t)$ increases with increasing $f_0$. Active-IMCT also predicts a similar result, as shown in Fig. \ref{sametau_activity}(g), with the inset showing overlapping $Q(t)$ for these chosen parameters. {The active IMCT theory does not predict the length scale directly, but it can be extracted from the peak height of the four-point susceptibility with the knowledge of the exponent relating to them as $\chi_4 \sim \xi^{2-\eta}$. Thus the estimate of the length scale
will be indirect. Within this caveat, the theory indeed suggests that the correlation length will increase with increasing activity keeping the relaxation time of the system the same.} Thus, the monotonic relation between the DH and the relaxation dynamics breaks down, suggesting a possible decoupling between the two. To show the generality of this striking result in active glasses, we have explored this behavior at different $\rho_a$ in our simulations. Figure \ref{sametau_activity}(h) shows the similar results for $\rho_a=1.0$, that is, $\chi_4^P$ grows as $f_0$ increases while $\tau_\alpha$ remains similar.
Similar conclusions are reached by studying the effects of activity on other dynamical quantities as well. Figure \ref{sametau_activity}(i) demonstrates that spatial correlation, computed by the excess part of the displacement-displacement correlation function, $\Gamma(r,\tau_\alpha)$, increases markedly with increasing activity, confirming the strong decoupling of DH and structural relaxation dynamics in active glasses. In {\color{blue}SM Sec.S2}, we have shown another striking difference between equilibrium glasses and active glasses. We have shown that DH increases with increasing fragility (see SM for definition) of liquids in equilibrium across model systems and spatial dimensions \cite{monojJPCB} whereas DH decreases sharply with increasing fragility in active glass. Although it is not immediately clear whether this behaviour is universal in active glasses but it is clear that marked deviation from equilibrium behaviour is solely due to non-equilibrium active forces.

\begin{figure}[!h]
\includegraphics[width=0.48\textwidth]{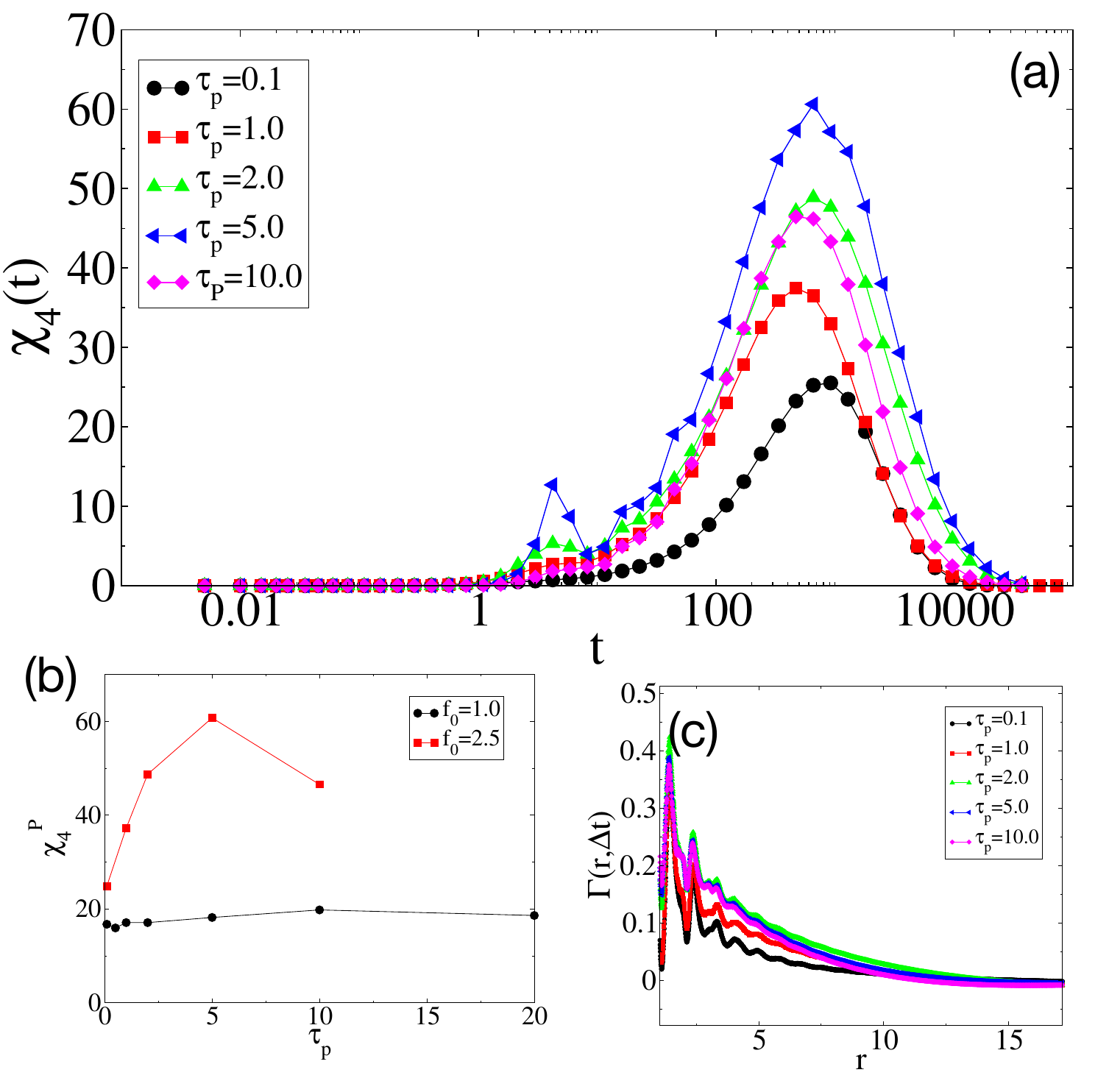}
\caption{(a)The time evolution of $\chi_4(t)$ for active system with different persistence time, and $f_0=2.5$. (b) The peak height for the case of $f_0=2.5$, increases with increasing $\tau_p$, and then decreases or tend to reach a saturation. However, system with $f0=1.0$, hardly shows weak $\tau_p$ dependence. (c) The excess displacement-displacement correlation function of $f_0=2.5$, also conveys the similar information.}
\label{TauP}
\end{figure}
{
A recent work \cite{KetaPRL} studied athermal over-damped Brownian particles interacting via repulsive potential and propelled by active Ornstein-Uhlenbeck force. They found that in the dense limit of highly persistent particles, the dynamic susceptibility $\chi_4(t)$ is significant at all times up to $\tau_\alpha$, interestingly it peaks with the caging parameter $a$ close to root-mean-squared-displacement, $a\sim \sqrt{\Delta r^2(t)}$. This picture of DH is very different from the passive supercooled liquids since the latter shows no significant small-time correlations. In this study, activity was varied by varying the persistence time, keeping the effective temperature the same. This is really interesting to see that these two very different systems show enhanced dynamic heterogeneity with increasing activity suggesting that the enhancement of dynamic heterogeneity due to active self-propulsion forces is a generic phenomenon and will be observed in a wide variety of active dense systems. To study the effect of variation of $\tau_p$ in our system we computed $\chi_4(t)$ at two different activity $f_0 = 1.0$ and $f_0 = 2.5$ and varied $\tau_p$ from $0.1$ to $20$ keeping effective temperature $\Teff$ same ({\it i.e.} $\tau_\alpha$ same). For low $f_0$ value, the effect of changing $\tau_p$ seems very minute, but at higher $f_0$, one sees a significant
change in peak height of $\chi_4(t)$. What is interesting is that dynamic heterogeneity increases monotonically with increasing persistence time up to a certain value and then starts to decrease or reaches saturation as shown in Fig.\ref{TauP}a for $f_0 = 2.5$. Fig.\ref{TauP}b shows the variation of $\chi_4^p$ as a function of $\tau_p$ for $f_0 = 1.0$ and $2.5$ respectively. Fig.\ref{TauP}c displacement-displacement correlation function for $f_0 = 2.5$ to highlight similar non-monotonic growth of DH with increasing $\tau_p$. These results are very interesting but we don't have a good microscopic understanding. Enhancement of DH at a small time scale has also been observed in the present model in \cite{SubhodeepSoftMatter2022} with changing $f_0$, the concentration ($c$) of active particles, but it is conjectured to be due to enhancement of long-range phone like excitation in the system at these time scale as opposed to the cooperative motion due to structural relaxation. Thus, it is certain that further detailed studies are required to explore other parts of the phase diagram as one can discover interesting physics in those regimes \cite{mandal2020}.
}

{
For completeness, it is important to discuss the existence of a few length scales that naturally appear
due to velocity correlation in these non-equilibrium systems. For example, one can define two
length scales related to bulk and shear modulus of the system as $\xi^2_T = \mu \tau_p/\zeta$ and
$\xi^2_L = (\mu +B) \tau_p/\zeta$ \cite{Henkes2020, Caprini2020PRL, Szamel_2021}, where $\mu$ and $B$ are the shear and bulk modulus of the system, and $\zeta$ is a damping coefficient of the effective medium where the particles are suspended. These length scales will be relevant at timescales shorter than $\tau_\alpha$, but at longer timescales studied in the present work, they may not play an important role, and both $\mu$ and $B$ in the liquid will be zero. Also, it is known that the long-range nature of velocity correlation in active systems will be present even at high temperatures where the dynamic heterogeneity as measured by $\chi_4(t)$ will be very small (nearly zero at high temperatures). $\xi_D$ in our system systematically grows with decreasing temperature along with the corresponding four-point susceptibility. Thus the length scale measured in our study seems to be purely due to supercooling effect and not due to the velocity correlation coming from the non-equilibrium nature of the active forcing.
}
\section*{Discussion}

We have studied the glassy dynamics and DH in model glass-forming systems in the presence of activity via a combination of simulations and an analytical theory. Although the relaxation dynamics of an active glass can be equilibrium-like at a $\Teff$, the notion that activity merely fluidizes the system with no other significant changes in its dynamical properties is not accurate. An active glass is fundamentally different from an equilibrium glass; the nontrivial effect of activity manifests in the DH. DH refers to the coexistence of fast and slow-moving regions, has emerged as one of the salient features of glassy systems \cite{ediger2000,giulioreview,berthier2011}, and {is believed to be responsible for non-Gaussian dynamical behaviour as well as break down the Stokes-Einstein relation between diffusion constant and viscosity of the liquid. However, active systems with the same relaxation time but with different degrees of activity can have varying DH and $\xi_D$. This masks the monotonic relation between DH and $\tau_\alpha$ in active glasses, suggests a decoupling between the two, and casts doubt on the presumed central role of DH in glassy dynamics. At this point, one can argue that increasing DH with increasing activity is not very surprising as systems with different activities are completely different systems, so it is probably not appropriate to compare them. It is important to highlight in this context that if the effective temperature is simply temperature-like, which the relaxation dynamics seem to suggest, then one expects a monotonic relation between the peak of $\chi_4(t)$ and the relaxation time. Often in biological systems, one school of thought is to assume activity as another parameter and not distinct systems. Although, even in equilibrium glasses, Ref. \cite{smarajitPNAS2009} has shown that the same length scale may not control the finite-size effects of DH and relaxation time, although their monotonic relation survives. Is it specific to active glasses? Whether the monotonic behavior in equilibrium glass is a coincidence or manifestation of something more fundamental remains an important open question. }

The nontrivial effects of activity on DH have significant implications for theories of glassy dynamics. Within the DH phenomenology, the glass transition is a critical phenomenon; $\chi_4^P$ and $\xi_D$ have power-law behaviors close to the transition point. Activity-dependence of the power-law exponents posits a significant challenge in developing a microscopic theory for such systems. Yet, our active-IMCT predicts the key characteristics of DH in active systems: the nontrivial behavior comes from an activity-specific term within the memory kernel, Eq. (\ref{sigmaterm}).
A striking effect of activity is the astonishingly large $\xi_D$ compared to that in passive systems. 
In particular, we show DH grows dramatically with increasing activity, even when $\tau_\alpha$ remains constant. Recent experiments on confluent cellular monolayers have found increased $\chi_4^P$ via the application of GTPase RAB5A, even when the system fluidizes, i.e., $\tau_\alpha$ decreases \cite{malinverno2017,cerbino2021}. Application of RAB5A can have several effects, such as modifying the junctional proteins, changing inter-cellular interaction \cite{souvik2021} and affecting the motor proteins \cite{palamidessi2019}. Via a combination of theory and experiment, Ref. \cite{giavazzi2017} has convincingly shown that at least one principal effect of RAB5A on the monolayer is higher cellular motility. Thus, these exciting results, though puzzling from the equilibrium glassy dynamics perspective that implies higher $\tau_\alpha$ at larger $\chi_4^P$, are entirely consistent in the light of our theory.
The theory developed here is independent of the microscopic details of a system, and hence, the results should be general and apply to a broad class of systems. The similar behavior in two distinct models in our simulations also supports this. Thus, our work provides a general framework to understand the effects of activity in various dense biological systems.

In conclusion, the dynamical heterogeneity in active glasses is inherently different from its equilibrium behaviour. Although relaxation dynamics in active systems can be equilibrium-like at a suitably defined $\Teff$, the DH has an entirely different behavior. Activity affects both the transition points and exponents of the power-law behaviors of both $\xi_D$ and $\chi_4^P$. In particular, systems with varying activity but the same $\tau_\alpha$ can have varying $\xi_D$; this highlights the decoupling of relaxation dynamics and DH, an effect that may have consequences in equilibrium systems \cite{smarajitPNAS2009}. Since the theory is quite general and agrees with different model active systems, we expect our results to apply to a broad class of biological systems.

\section*{Materials and Methods}
The Kob-Andersen binary mixture with 80\% A-type and 20\% B-type particles interacting via the 
Lennard-Jones pair potential,
\begin{equation}
\Phi\left( \f{r_{ij}}{\sigma_{ij}} \right)=4\epsilon_{ij}\left[ \left(\f{\sigma_{ij}}{r_{ij}}\right)^{12}-\left( \f{\sigma_{ij}}{r_{ij}} \right)^6 \right],
\end{equation}
where $r$ is the distance between two particles and the indices $i$ and $j$ can be A or B. The values of $\sigma_{ij}$ and $\epsilon_{ij}$ are chosen to be: $\sigma_{AB} = 0.8 \sigma_{AA}$, $\sigma_{BB} = 0.88 \sigma_{AA}$, $\epsilon_{AB} = 1.5\epsilon{AA}$, and $\epsilon_{AB} = 0.5\epsilon_{AA}$. We set a cutoff in the potential at $r_{ij} = 2.5\sigma_{ij}$ and shift it accordingly. We set the unit of length and energy as $\sigma_{AA} = 1$ and $\epsilon_{AA}= 1$ and fix the overall density $\rho$ at 1.2. We use a quadratic polynomial to make the potential and its first two derivatives smooth at the cutoff distance. We chose another model that interpolates between finite-$T$ glasses and hard-sphere glasses and has been studied extensively in the context of jamming physics. This is a $50 : 50$ binary mixture with a diameter ratio of $1.4$. 
In this work, we keep $\tau_p=1.0$ fixed, and both models of activity, discussed in Ref. \cite{activerfot}, are then identical. {We performed most of our simulations with system size $N = 50,000$ particle, but to check the finite size effects especially at low temperatures and higher activities we have performed additional simulations with $N = 100,000$ particles.}

{We introduced activity for a fraction $\rho_a$ of the total number of particles in the system. These active particles are chosen randomly and 
assigned a self-propulsion force of the form $\vec{f}= f_0(k_x\hat{x} + k_y\hat{y} + k_z\hat{z})$, where $k_x$, $k_y$, $k_z$ are $\pm1$, 
chosen randomly to maintain the momentum conservation. After every persistence time $\tau_p$, the set of  values of $k_x$, $k_y$, $k_z$ are changed, maintaining the momentum conservation. In this work, we mostly keep $\rho_a= 0.1$ and always use $\tau_p= 1.0$ and study the effect of activity as a function of $f_0$ only. We have integrated the Newton's equations of motion in  a constant particle number ($N$), volume ($V$) and temperature ($T$) (NVT) ensemble. We have used Gaussian operator-splitting  \cite{zhang1997} and Nose-Hoover thermostat in our simulation to keep kinetic temperature, $T$ constant. It is defined as  $T=\frac{2}{3N-3}\sum_{i=1}^N\frac{1}{2}\vec{v}_i.\vec{v}_i$}

\section*{Dynamical quantities}
To characterise the dynamics of glassy systems under various active forcing, $f_0$, we measure the following dynamical quantities.

\subsection*{Mean-squared displacement}
The mean-squared displacement (MSD) $\left< \Delta r(t)^2 \right>$ is defined as
\begin{equation}
\left< \Delta r(t)^2 \right> = \left< \frac{1}{N}\sum_{i=1}^{N}\mid{\textbf{r}_i(t+t_0) - \textbf{r}_i(t_0)}\mid^2 \right>.
\end{equation}

\subsection*{Overlap correlation function}
The two point overlap correlation function $Q(t)$ is defined as
\begin{equation}
Q(t) =\langle\tilde{Q}(t)\rangle= \left< \frac{1}{N}\sum_{i=1}^{N}w(\mid{\textbf{r}_i(t+t_0) - \textbf{r}_i(t_0)}\mid) \right>,
\end{equation}

\begin{equation}
\text{where,} \,\,\, w(r) = \begin{cases}1, \quad  \text{if $r \leq a$}\\
0,  \quad \text{otherwise},
\end{cases}
\end{equation}
$\left< ... \right>$ represents the average over time origin $t_0$ and 10 independently generated configurations. $N$ is the total number of particles and the parameter $a$ is associated with the typical vibrational amplitude of the caged particles. We have 
used the parameter $a = 0.3$ in our analyses and verified that a moderate variation in $a$ does not affect our results.

\subsection*{Four-point dynamical susceptibility}
The four-point dynamical susceptibility, $\chi_4(t)$, is defined in terms of the fluctuations of the two point overlap correlation function as
\begin{equation}
\chi_4(t) = N[\left<\tilde{Q}^2(t)\right> - \left<\tilde{Q}(t)\right>^2]
\end{equation}

\subsection*{Excess displacement correlation function}
The excess displacement correlation $\Gamma(r,\Delta t)$ defined as
\begin{equation}
\Gamma(r,\Delta t) = \frac{g_{uu}(r,\Delta t)}{g(r)} - 1, 
\end{equation}
where $g(r)$ is the radial pair correlation function, defined as
\begin{equation}
g(r) = \frac{1}{\rho N} \left<\sum_{i,j=1,j\neq i}^{N} \delta(\textbf{r} + \textbf{r}_i(0) - \textbf{r}_j(0))\right>,
\end{equation}
and the spatial correlation of the particle displacements, $g_{uu}(r, \Delta t)$, is 
\begin{equation}
g_{uu}(r,\Delta t) = \frac{\left<\sum_{i,j=1,j\neq i}^{N}u_i(t,\Delta t)u_j(t,\Delta t)\delta(r-\mid\textbf{r}_{ij}(t)\mid)\right>}{4\pi r^2\Delta rN\rho\left<u(\Delta t)\right>^2},
\end{equation}
where $u_i(t,\Delta t) = \mid\textbf{r}_i(t+\Delta t) - \textbf{r}_i(t)\mid$ is the scalar displacement of the particle between time $t$ and $t+\Delta t$.

\begin{figure*}[htpb]
	\begin{center}
		\includegraphics[width=0.95\textwidth]{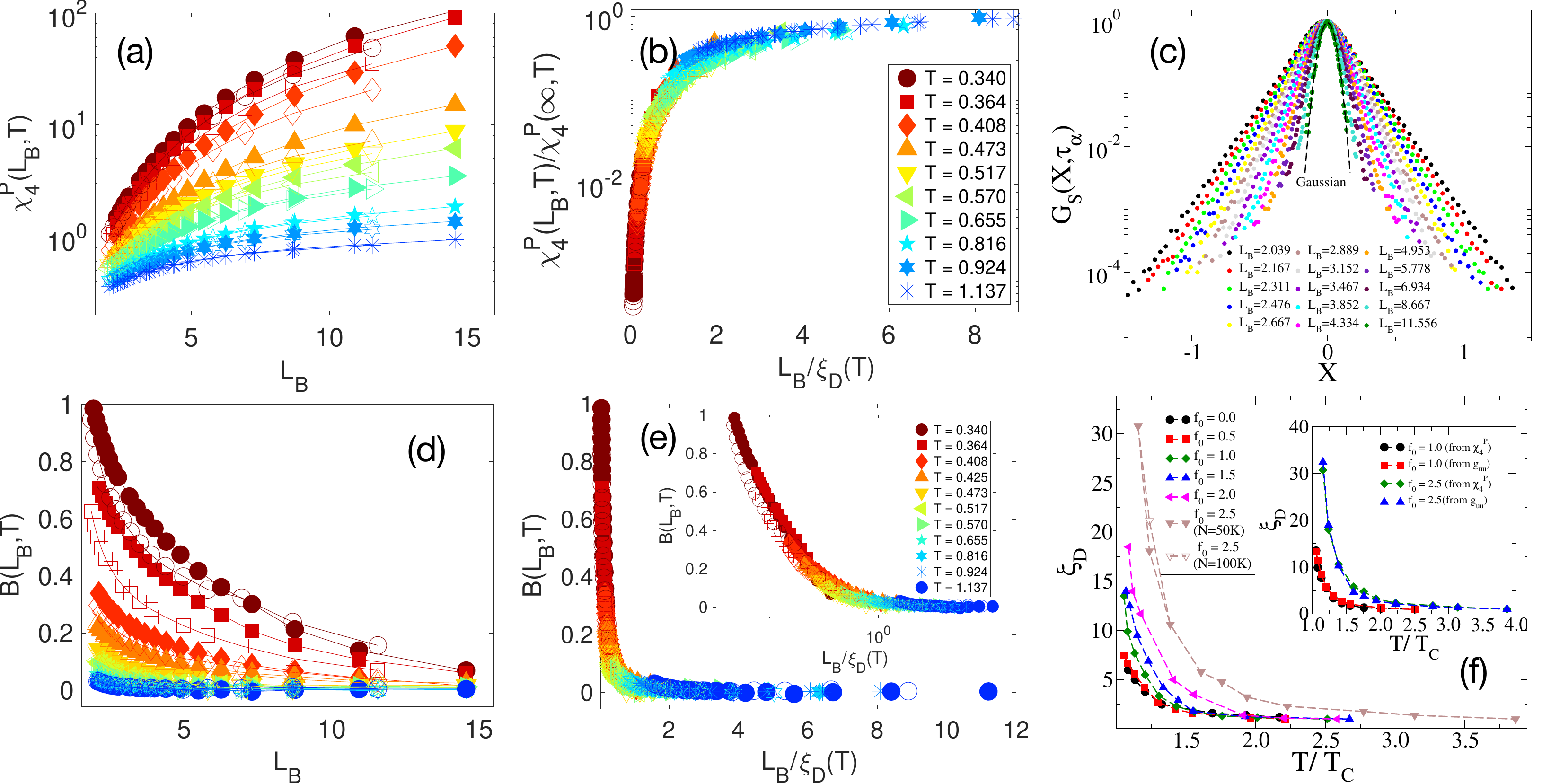}
		\caption{{\bf Estimation of dynamical heterogeneity length scale.} (a) The block size-dependence of the peak height of $\chi_4(t)$, $\chi_4^P$, for an active system with $f_0 = 2.5$, and at different $T$ [legend is same as in (b)]. In these plots, open symbols are for $N=50,000$ system size, while the filled symbols are for $N=100,000$ system.	
			(b) The finite-size scaling collapse of the data in panel (a) to obtain the dynamic heterogeneity length scale, $\xi_D$ (Eq. \ref{blockanalysis}). (c) Block size-dependence of the self part of the van Hove correlation function, $G_s(x,\tau_\alpha)$, computed at structural relaxation time. $G_s(x,\tau_\alpha)$ shows non-Gaussian behaviour at small $L_B$ and becomes progressively Gaussian with increasing $L_B$.  $G_s(x,\tau_\alpha)$ becomes Gaussian when $L_B$ exceeds a crossover length scale that gives $\xi_D$.
			(d) Binder cumulant of the block averaged $G_s(x,\tau_\alpha)$ at different $T$. (e) Scaling collapse of the Binder cumulant computed in panel (d) to obtain $\xi_D$. {Inset contains the same data collapse but in log scale.} (f) $\xi_D$ as a function of scaled temperature, $T/T_C$, where $T_C$ is the MCT transition temperature at various activities. Note that the same length scale collapses both the $\chi_4^P$ and Binder cumulant of van Hove function data.  {\bf Inset:}  Comparison of different estimates of $\xi_D$, obtained via block analysis of $\chi_4^P$ and using 
			spatial displacements correlation function, $g_{uu}(r,\tau_\alpha)$ (see text for details). The excellent agreement suggests the robustness of the methods and the reliability of the estimates of $\xi_D$. {The main panel also contains the comparison of $\xi_D$ for different system size for $f_0=2.5$ case.}  }
		\vskip -0.1in
		\label{lengthscale}
	\end{center}
\end{figure*}
\subsection*{van Hove correlation function}
The self part of the van Hove correlation function is defined as
\begin{equation}
G_s (x,\tau) = \left< \delta [x-(x_i(\tau)-x_i(0))] \right>
\end{equation}

\subsection{Computation of the dynamical heterogeneity length scale}
\label{comp_xi}

As we discussed in the main text, we have employed four distinct methods to reliably compute the dynamic length scale, $\xi_D$, associated with DH itself. We now briefly discuss these methods and refer the reader to the SM for the technical details.

First, in the block analysis method, a somewhat larger system with linear size $L$ is simulated at the desired state points; the system is then divided into smaller blocks of length $L_B = L/n$, where $n$ is the number of blocks chosen, as illustrated in Fig. \ref{lengthscale}(a). The largest blocks in our analyses have a linear size $L/3$ to minimize other boundary effects in the results. 
{The dynamic susceptibility is defined as,
\begin{equation}
\chi_4(L_B,t) = \frac{NL_B^3}{L^3}\left<[Q^i(L_B,t) - \left<Q(L_B,t)\right>]^2\right>,
\end{equation}
where $N$ is the total number of particles in the system, and $N_B$ are the number of blocks with size $L_B$.Here, $\left< ... \right>$ represents the average over time origin $t_0$, and $N_B$ number of blocks. Then the obtained $\chi_4(L_B,t)$, is averaged over 10 ensembles. The overlap for each block is defined as, 
\begin{equation}
Q^i(L_B,t) = \frac{1}{n_i}\sum_{j=1}^{n_i} \left< w(\mid{\textbf{r}_j(t) - \textbf{r}_j(0)}\mid) \right>,
\end{equation}
with $n_i$ being the number of particles in the $i^{th}$ block at time $t = 0$, and the window function $w(x) = \Theta(a-x)$, where $\Theta$ is the Heaviside step function, the parameter $a$ is chosen to remove the decorrelation from vibrations of particles inside the cages formed by their neighbours. }The dynamic length scale is obtained from a detailed finite-size scaling (FSS) analysis of the dependence of the four-point dynamic susceptibility on the block size by assuming the following scaling ansatz,
\begin{equation}\label{blockanalysis}
\chi_4^P(L_B,T) = \chi_0(T)\mathcal{F}\left(\frac{L_B}{\xi_D}\right)
\end{equation}
where, $\chi_0(T) = \lim_{L_B\to\infty}\chi_4^P(L_B,T)$. The data for all temperatures can be collapsed to a master curve using the two parameters, $\chi_4^P(\infty,T)$ and $\xi_D$.

{Figure \ref{lengthscale}(a) shows $\chi_4^P(L_B,T)$ as a function of $L_B$ at various $T$ for the $f_0 = 2.5$. The filled symbols are for system size $N = 100,000$ particles and open symbols are for system size $N = 50,000$ particles. Fig. ~\ref{lengthscale}(b) shows the data collapse, Eq. (\ref{blockanalysis}), to obtain $\xi_D$, as shown in Fig.~\ref{lengthscale}(f). Note that for lower temperatures and larger activities the correlation length is close to half of the linear size of the simulations box which might raise concerns about the possible finite size effects in them. If one notice the data of $\chi_4^P(L_B,T)$ for $N = 50,000$ and $N = 100,000$ system sizes (open and solid symbols respectively), then one do observe some finite size effects but as the length scale is obtained using how $\chi_4^P(L_B,T)$ depends on $L_B$, rather than their absolute values, the obtained dynamic length scale, $\xi_D$ does not have strong finite size effects as the data from both the system sizes can be collapsed on a single master curve using the same length scale but by adjusting $\chi_4^P(\infty,T)$ slightly.} The analyses for all other values of $f_0$ are reported in SM along with other relevant discussions. The excellent data collapses observed in the analyses suggest the reliability of the extracted $\xi_D$. However, in the absence of previous data to compare with, we reconfirm the reliability of the estimated $\xi_D$ via three other methods as mentioned earlier.

Second, we have used the non-Gaussian nature of particle displacements at times  comparable to the relaxation time in a glassy system to estimate $\xi$. Reference \cite{bhanu2018} has demonstrated that $\xi_D$ can be obtained by analyzing the non-Gaussian behaviour of the self part of the van Hove function, $G_s(x,\tau_\alpha)$ (defined in Method section and elaborated in SM). 
Intuitively, as the coarse-graining volume gradually increases, the system eventually becomes spatially homogeneous once $L_B$ exceeds the typical $\xi_D$. Figure~\ref{lengthscale}(c) shows $G_s(x,\tau_\alpha)$ with increasing $L_B$ illustrating the change of nature of $G_s(x,\tau_\alpha)$, from non-Gaussian to Gaussian, as $L_B$ increases. $G_s(x,\tau_\alpha)$ becomes Gaussian at a crossover scale $L_B$, which is different at each $T$; the $T$-dependence of this crossover scale has the same $T$-dependence of $\xi_D$ \cite{bhanu2018}. For an unbiased estimate of this crossover length scale, we compute the Binder cumulant of the distribution as shown in Fig.~\ref{lengthscale}(d); increase of the Binder cumulant with a decrease in $L_B$ becomes higher at lower $T$.
As $L_B \to \infty$, Binder cumulant should reach 0 at any $T$. We can do a one-parameter scaling collapse (Fig.~\ref{lengthscale}e) using the scaling ansatz $B(L_B,T) = \mathcal{G}\left( L_B/\xi_D\right)$ where $\xi_D$ is the crossover length scale, same as the one used in collapsing the $\chi_4^P(L_B,T)$ data. {To check finite size effects, we  have done analysis for $N = 100,000$ and $N = 50,000$ particles as shown via filled and open symbols respectively. One do see some amount of finite size effects but the obtained length scale has not been affected significantly. In the inset of Fig.~\ref{lengthscale}e we show the same data collapse but in log-log scale for better clarity.}
The quality of the data collapse shows that these two methods yield the same $\xi_D$. {In Fig.~\ref{lengthscale}f main panel, we show the growth of $\xi_D$ for all activities and especially for $f_0 = 2.5$ activity we show the results using two different system size. They are found to be very similar, confirming once more that our estimation of $\xi_D$ is not affected by finite size effects severely.}

Third, we have extracted the same length scale through studying the spatial correlation of the displacement fields of the particles over the structural relaxation time scale, characterized by the correlation function $g_{uu}(r,\tau_\alpha)$, where $r$ is the radial distance ({see the definition in the Method section and further details in SM}) \cite{indrajit2020}. In the inset of Fig. \ref{lengthscale}(f), we show the comparison of the length scales; the excellent agreement again highlights the reliability of the obtained length scale. Figure~\ref{lengthscale}(f) shows the growth of $\xi_D$ for various activities as a function of $T/T_C$, where $T_C$ is the MCT transition temperature ({\color{blue}see SM, Sec. 3}). Note the dramatic growth of $\xi_D$ with increasing activity, this is one of the main results of this work, and we now discuss the salient features of this observation.

Finally, we have studied the spatial correlation of the four-point function to extract the growing length scales $\xi_d$ of the dynamic heterogeneity. In this study, we utilize the frequently used four-point structure factor using the overlap function $Q$, which is defined as
\begin{equation}
S_4(q,\tau_\alpha) = \frac{1}{N} \left<Q(q,t)Q(-q,t)\right>
\end{equation}
The behavior of $S_4(q,\tau_\alpha)$ at small wave-numbers can be described by the Ornstein–Zernike (OZ) form
\begin{equation}
S_4(q,\tau_\alpha) = \frac{S_4(q=0,\tau_\alpha)}{1+(q\xi_d)^2}
\end{equation}

\begin{figure}
	\centering
	\includegraphics[width=8.6cm]{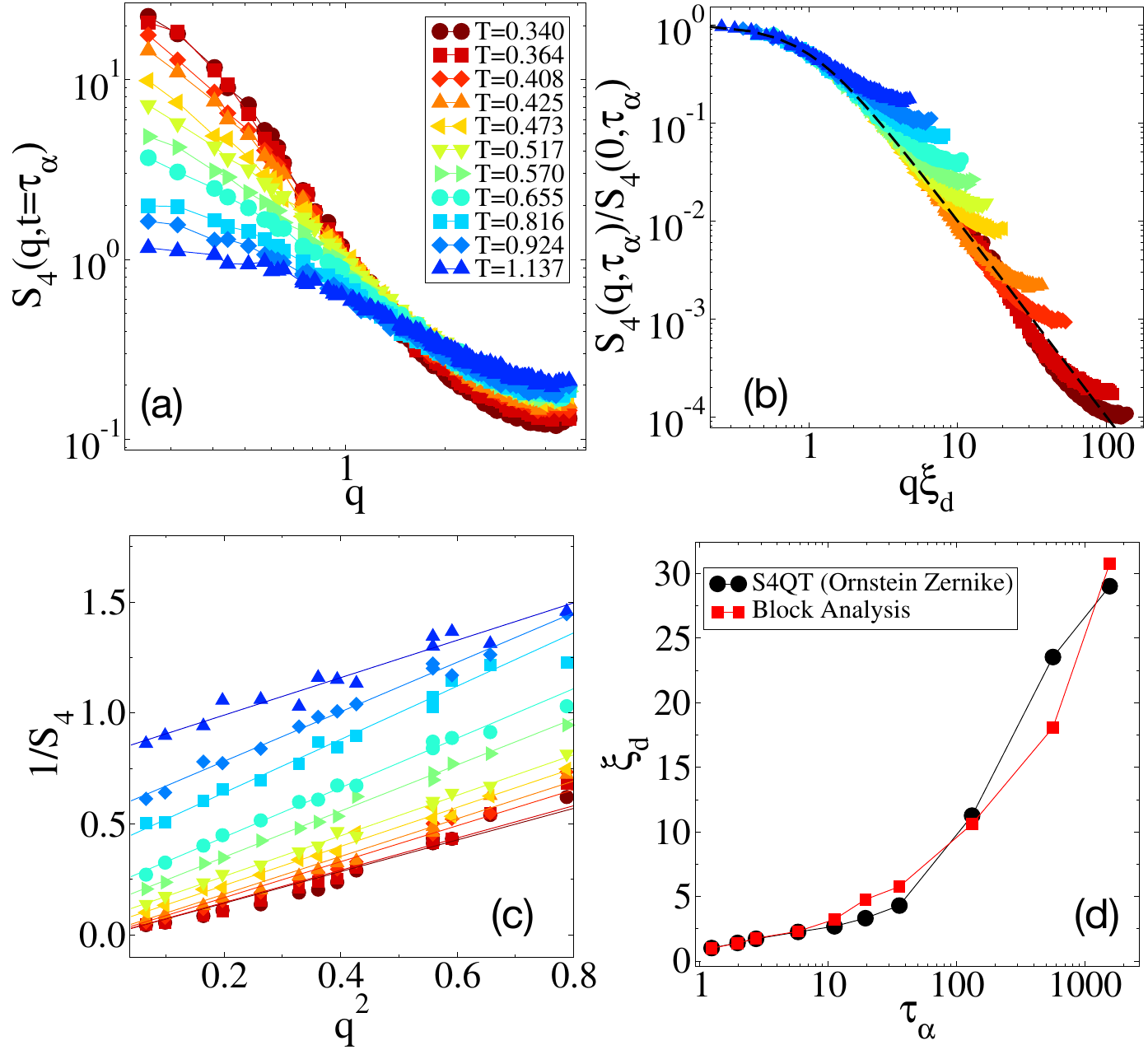}
	\caption{(a)$S_4(q,\tau_\alpha)$ for the $f_0 = 2.5$ as a function of $q$ at different temperatures. (b)The scaling collapse using the dynamic length scale $\xi_d$. (c)$1/S_4(q,\tau_\alpha)$ for the $f_0 = 2.5$ as a function of $q^2$ from which we obtained the dynamical length scale $\xi_d$. (d) Comparison of $\xi_d$ obtained from $S_4(q,\tau_\alpha)$ and from block analysis}
	\label{s4qt}
\end{figure}

In Fig.\ref{s4qt}(a) we show the wave-number dependence of $S_4(q,t)$ on the time scale $\tau_\alpha$ at different temperatures for $f_0 = 2.5$. Fig.\ref{s4qt}(b) displays
the scaled function $S_4(q,t=\tau_\alpha)/S_4(q=0,t=\tau_\alpha)$ as a function of $q\xi_d(\tau_\alpha)$. At lower $q$ limit one can also compute the $\xi_d$ directly following the Ornstein–Zernike relation. {The dotted line in Fig.\ref{s4qt}(b) is simply $f(x) = 1/(1+x^2)$. This also shows the robustness of the OZ theory prediction even for non-equilibrium systems.} In Fig.\ref{s4qt}(c) we show $1/S_4(q,\tau_\alpha)$ as a function of $q^2$. The continuous lines are the linear fit of the data. Hence we compute the $\xi_d$ from the fitted parameters. In Fig.\ref{s4qt}(d) we show the comparison of dynamic length scale obtained from the analysis of $S_4(q,t)$ and obtained from the finite size scaling of $\chi_4$.

{
\subsection{Brownian dynamics}
\label{Brown}
\begin{figure}
\includegraphics[width=0.48\textwidth]{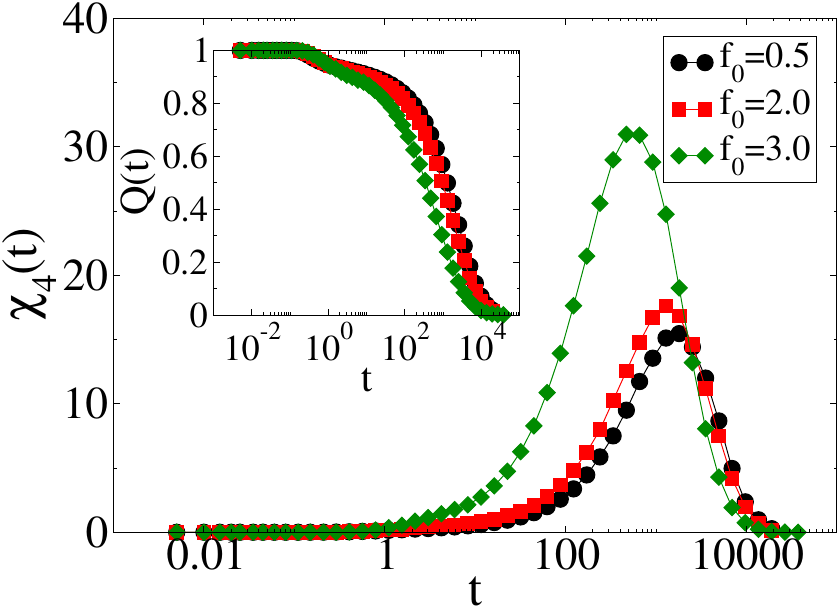}
\caption{The time evolution of $\chi_4(t)$ obtained via Brownian dynamics is shown for systems with different activities and similar relaxation times. \textbf{Inset:} $Q(t)$ is plotted for the same set of parameters.}
\label{BrownFig}
\end{figure}
To study the effect of the details of the dynamics, we performed Brownian dynamics simulations using the predictor-corrector algorithm \cite{Gleim1998} and set the self-diffusion constant $D_0$ to unity. We kept the nature of the activity the same in the simulation as in Molecular Dynamics simulations. We confirm again that our findings that DH increases with activity are robust. In Fig.\ref{BrownFig}, we display the increase in peak height of $\chi_4(t)$ for systems with similar $\tau_\alpha$, as activity increases. We also note here that a detailed study of the effect of the inertia of the particles will be very important to have a better understanding of the systems; for example, how our results change with the systematic tuning of the particle's inertia will be very interesting as recent studies seem to suggest \cite{Vrugt2023} very rich behaviour.}

\vskip +0.2in
\noindent{\bf Author Contributions:} SK designed the research project. KP and AM performed the 
research. SKN developed the active MCT formalism. AM, KP, SKN and SK analysed the data. AM performed additional simulations with larger system size and 
produced Brownian dynamics results. SKN and SK wrote the paper with inputs from KP and AM.

\vskip +0.2in
\noindent{\bf Acknowledgments:} We thank R. Cerbino, F. Glavazzi, G. Scita, Indrajit Tah and Rituparno Mandal for fruitful discussions. 
We acknowledge support of the Department of Atomic Energy, Government of India, under Project Identification No. RTI 4007. 
SK acknowledges support from Core Research Grant CRG/2019/005373 from Science and Engineering Research Board (SERB) as 
well as Swarna Jayanti Fellowship grants DST/SJF/PSA-01/2018-19 and SB/SFJ/2019-20/05. KP acknowledges financial support 
from SB/SFJ/2019-20/05. SKN thanks SERB for grant via SRG/2021/002014.


\bibliography{dynamiclength_MCT_ref}

\end{document}


\title{Dynamical heterogeneity in active glasses is inherently different from its equilibrium behavior - Supplementary Materials}

	
\author{Kallol Paul}
\thanks{Contributed equally}
	\affiliation{TIFR Center for Interdisciplinary Science, Tata Institute of Fundamental Research, 36/P Gopanpally Village, Serilingampally Mandal, RR District, Hyderabad, 500046, Telangana, India}	
	\author{Anoop Mutneja}
	\thanks{Contributed equally}
	\affiliation{TIFR Center for Interdisciplinary Science, Tata Institute of Fundamental Research, 36/P Gopanpally Village, Serilingampally Mandal, RR District, Hyderabad, 500046, Telangana, India}
	\author{Saroj Kumar Nandi} 
	\affiliation{TIFR Center for Interdisciplinary Science, Tata Institute of Fundamental Research, 36/P Gopanpally Village, Serilingampally Mandal, RR District, Hyderabad, 500046, Telangana, India}
	\author{Smarajit Karmakar}
	\email{smarajit@tifrh.res.in}
	\affiliation{TIFR Center for Interdisciplinary Science, Tata Institute of Fundamental Research, 36/P Gopanpally Village, Serilingampally Mandal, RR District, Hyderabad, 500046, Telangana, India}
\maketitle


\section{Computing Dynamic Heterogeneity Length Scale}
We have used three different techniques to compute the dynamic heterogeneity length scale $\xi_D$ for all studied activities. 

\subsection{Block analysis of $\chi_4^P$}
The length scale 
associated with dynamical heterogeneity is obtained from a finite-size scaling analysis of the dependence of the four-point dynamic susceptibility,
$\chi_4(t)$ (see definition later), on the block size, $L_B$ as discussed in the main article. All the simulations are carried out for moderately 
large system size, $N = 50000$. We checked possible finite size effects at low temperatures and at higher activities by doing simulation of $N = 100000$
particles. Below we have repeated some of the definitions that are already mentioned in the main article for completeness. We then construct blocks of size $L_B = L/n$, where $n \in \{3, 4, 5, ...\}$ and calculate various dynamic quantities using the particles which are 
present inside one such box at a chosen time origin. Then we compute the self overlap correlation function, $Q(L_B,t)$, at a particular block size,
\begin{equation}
Q(L_B,t) = \frac{1}{N_B}\sum_{i=1}^{N_B}\frac{1}{n_i}\sum_{j=1}^{n_i} \left< w(\mid{\textbf{r}_j(t) - \textbf{r}_j(0)}\mid) \right>,
\end{equation}
where $N_B$ is the number of blocks with size $L_B$, $n_i$, the number of particles in the $i$th block at time $t = 0$. $w(x) = \Theta(a-x)$ is the window function, where $\Theta$ is the Heaviside step function. The value of the parameter
$a$ is chosen to remove the de-correlation arising from vibrations of particles inside the cages formed by their neighbours. The dynamical 
susceptibility associated with blocks of size $L_B$ is then defined as follows
\begin{equation}
\chi_4(L_B,t) = \frac{NL_B^3}{L_0^3}\left<[Q(L_B,t) - \left<Q(L_B,t)\right>]^2\right>
\end{equation}

\begin{figure*}[!h]
\begin{center}
\includegraphics[width=1.\textwidth]{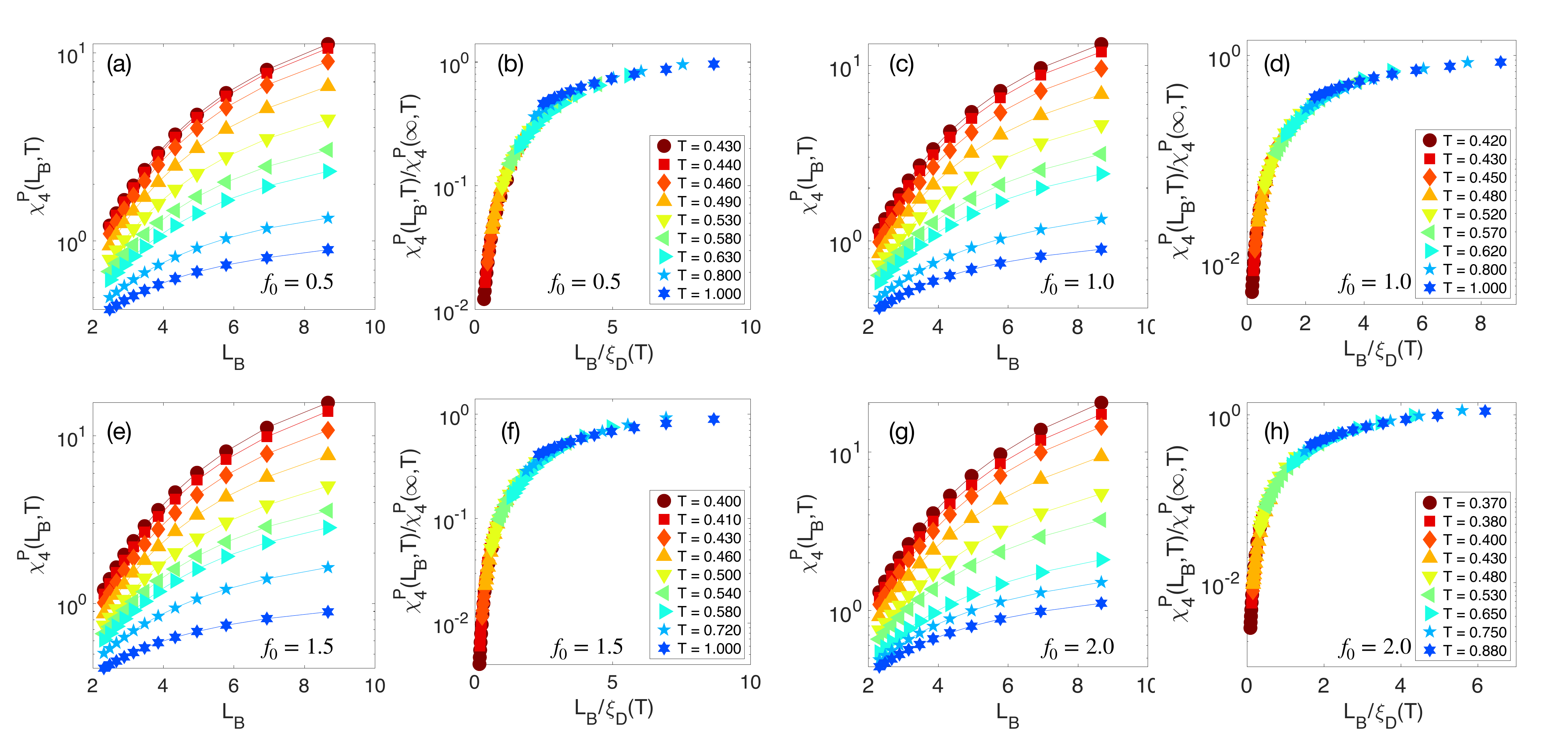}
\caption{Block size dependence of $\chi_4^P$ and finite size scaling for activity $f_0$ = 0.5 (a \& b), $f_0$ = 1.0 (c \& (d), $f_0$ = 1.5 (e \& f), 
$f_0$ = 2.0 (g \& h)}
\label{chi4p}
\end{center}
\end{figure*}

\begin{figure*}[!h]
\begin{center}
\includegraphics[width=1.\textwidth]{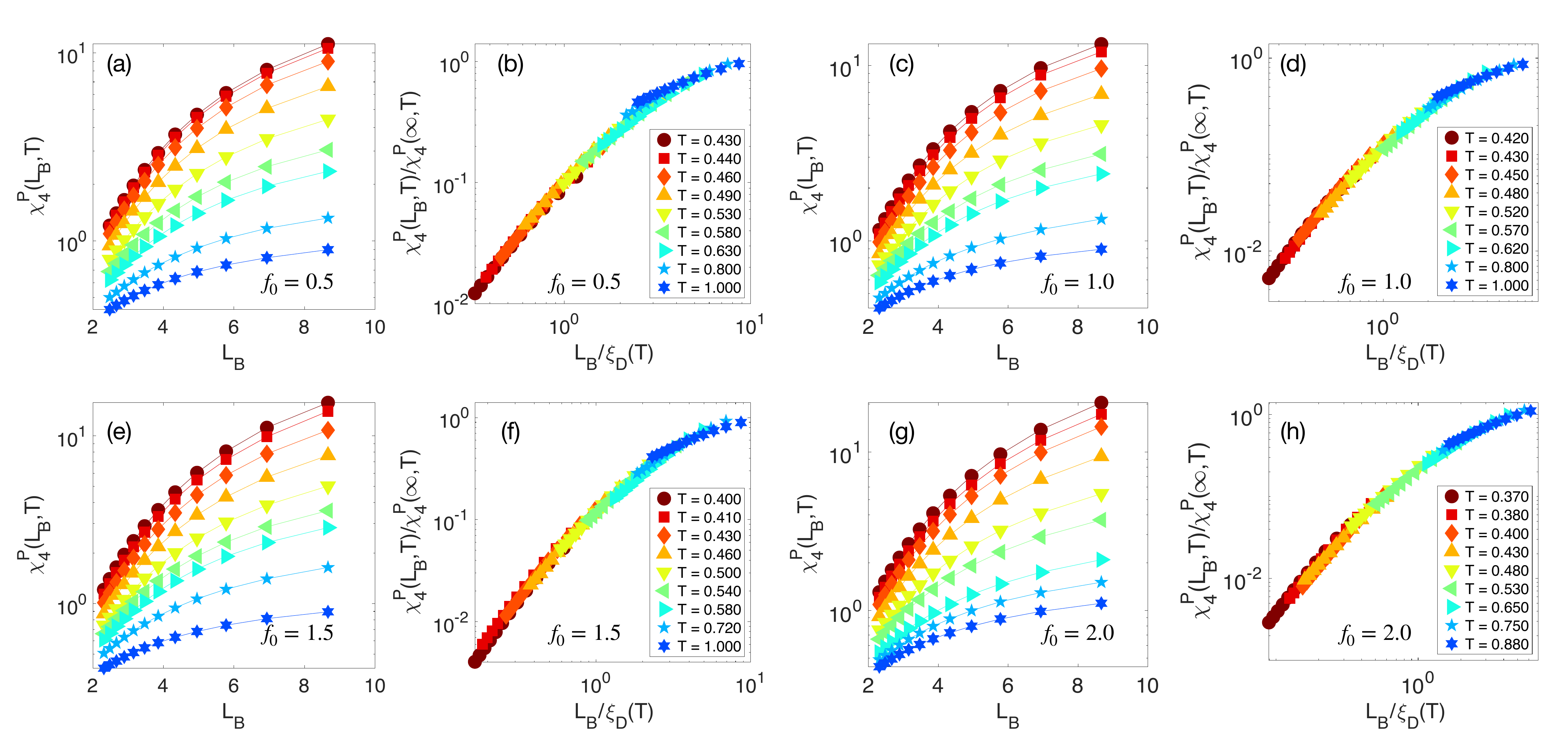}
\caption{Block size dependence of $\chi_4^P$ and finite size scaling for activity $f_0$ = 0.5 (a \& b), $f_0$ = 1.0 (c \& (d), $f_0$ = 1.5 (e \& f), 
$f_0$ = 2.0 (g \& h)}
\label{chi4p}
\end{center}
\end{figure*}

For each $f_0$, we consider the dependence of $\chi_4^P(L_B,T)$, the peak value of $\chi_4(L_B,t)$ at temperature $T$ , on the block size 
$L_B$ for a fixed value of $N = \rho L^3$. This dependence is shown in Fig.\ref{chi4p} The left panel of each sub figure shows the data for 
$\chi_4^P(L_B,T)$ as a function of the block length $L_B$ for different $T$. The peak value of the dynamical susceptibility at a given 
$T$ grows with $L_B$ and saturates at a $T$-dependent value $\chi_4^P(\infty,T)$.
The dependence of $\chi_4^P(L_B,T)$ on $L_B$ is expected to exhibit the following Finite Size Scaling (FSS) form:
\begin{equation}
\chi_4^P(L_B,T) = \chi_0(T)\mathcal{F}\left(\frac{L_B}{\xi}\right)
\end{equation}
where,
\begin{equation}
\chi_0(T) = \lim_{L_B\to\infty}\chi_4^P(L_B,T)
\end{equation}
and $\xi$ is a characteristic scaling length scale. The data for all temperatures can be collapsed to a master curve
using the two parameters, $\chi_4^P(\infty,T)$ and $\xi$, at each temperature, as shown in Fig. \ref{chi4p} for different activities.
This length scale is found to be the same as the dynamical heterogeneity length scale, $\xi_D$, as shown in \cite{saurish2017} for the equilibrium system.
The excellent data collapse confirms that the extracted length scales will be very reliable with small error bars (smaller than the point sizes in our plots).

\subsection{Block analysis of van Hove function}
The distribution of particle-displacements, known as the van Hove function, shows non-Gaussian behavior with exponential tail in the supercooled regime in glass-forming liquids. The non-Gaussian nature can be understood in terms of spatial and temporal heterogeneous dynamics and the exponential tail is a manifestation of dynamic heterogeneity. For completeness and ease of reading here also we repeat some of the definitions that are already mentioned in the main article. The van Hove correlation function is 
formally defined as 
\begin{equation}
G_s (x,\tau) = \langle \delta [x - (x_i(\tau) - x_i(0))] \rangle,
\end{equation}
where the $\langle\cdots\rangle$ implies the averaging over different statistically independent samples as well as the time origin averaging. 
We have performed systematic spatial coarse-graining of the dynamics to extract $\xi_D$ of the system as demonstrated in \cite{bhanu2018}. 
To do that we have used the method of block analysis where the whole simulation box is divided into smaller blocks of length, $L_B$. Thus 
for a block size of $L_B$, the number of particles in that block will be $N_B = \rho L_B^3$. Now we have defined a coarse-grained displacement 
of the particles in the $j^{th}$ block as
\begin{equation}
\Delta x_j^B(\tau) = \frac{1}{n_j} \sum_{i=1}^{n_j} [x_i(\tau)-x_i(0)],
\end{equation}
where $n_j$ is the number of particles in the $j^{th}$ block. Note that this number can be different for different blocks. Then we have defined the 
van Hove function for block as
\begin{equation}
G_s^B (x,\tau_\alpha) = \left<\frac{1}{N_B} \sum_{j=1}^{N_B} \delta[x-\Delta x_j^B(\tau_\alpha)]\right>.
\end{equation}
\begin{figure*}[!h]
\begin{center}
\includegraphics[width=0.9\textwidth]{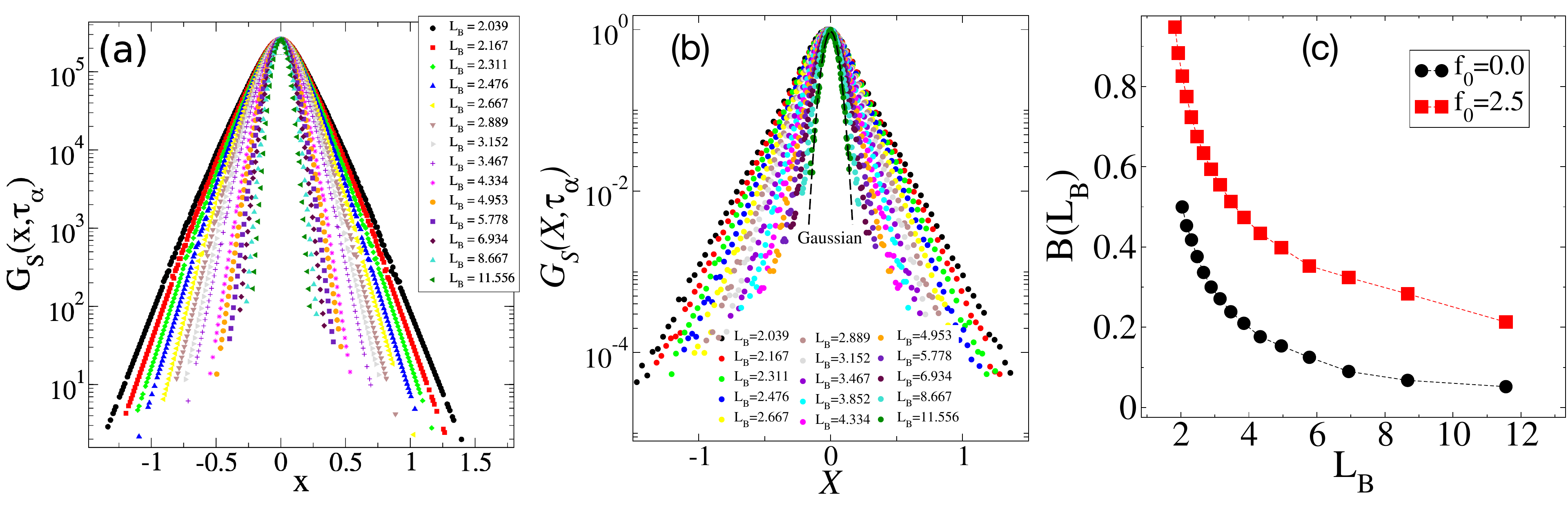}
\caption{The van Hove correlation function of different block lengths at lowest temperature for $f_0= 0.0$ (a), $f_0 = 2.5$ (b), and corresponding Binder Cumulants (c).}
\label{vanHove}
\end{center}
%
\begin{center}
\vskip +0.1in
\includegraphics[width=1.\textwidth]{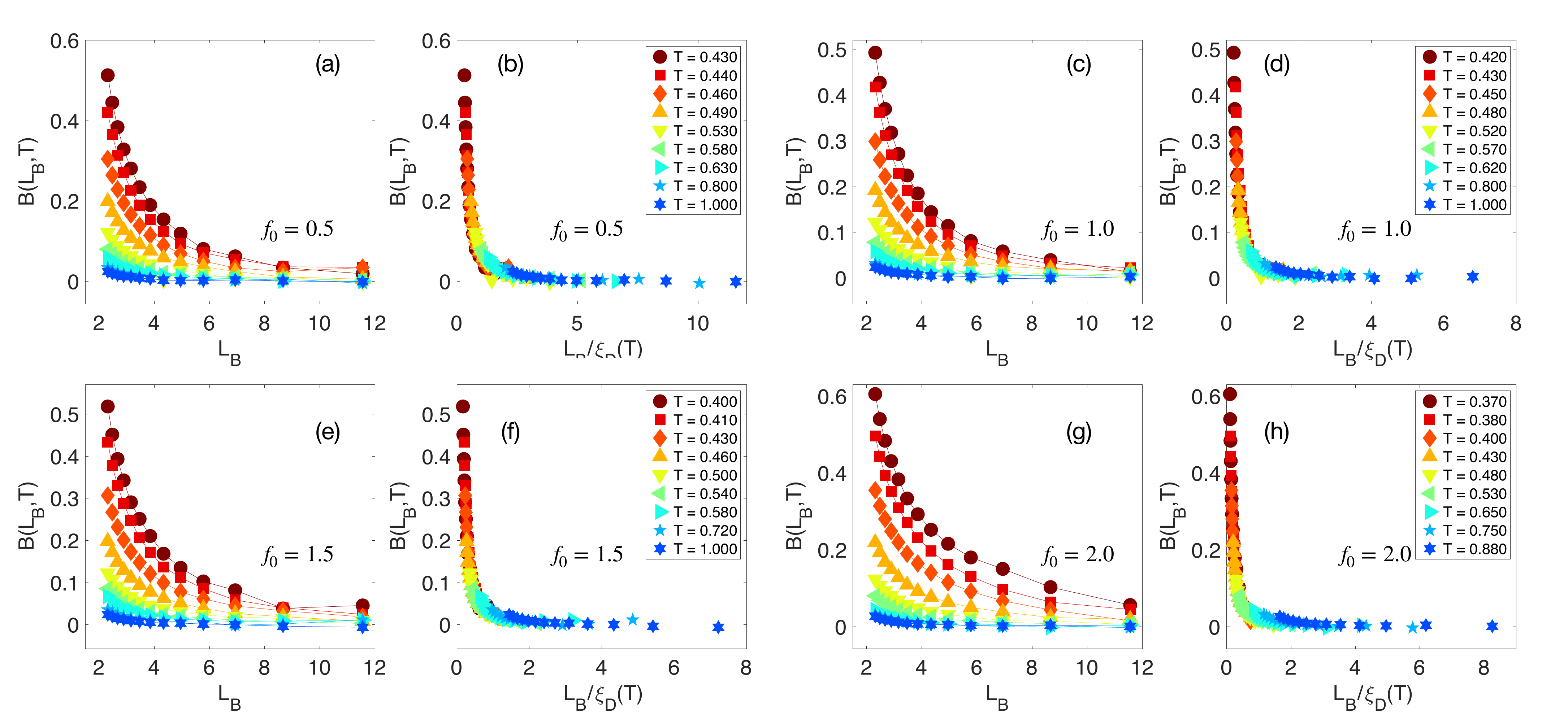}
\caption{Block size dependence of Binder Cumulant of Van-Hove function and finite size scaling for activity $f_0$ = 0.5 (a) and (b), $f_0$ = 1.0 (c) and (d), $f_0$ = 1.5 (e) and (f), $f_0$ = 2.0 (g) and (h).}
\label{binder}
\end{center}
\end{figure*}

By increasing the block length $L_B$, we have then studied the non-Gaussianity for different block lengths. One can observe that the 
non-Gaussianity increases with decreasing block size in Fig. \ref{vanHove}a (for $f_0 = 0.0$) and Fig. \ref{vanHove}b (for $f_0 = 2.5$). 
To measure the non-Gaussianity, we have calculated the Binder Cumulant of the distribution, which is defined as 
\begin{equation}
B(L_B,T) = 1 - \frac{\left<x^4\right>}{3\left<x^2\right>^2}.
\end{equation}
To extract the temperature dependence of the coarse-graining length scale, we performed the finite-size scaling analysis of the Binder 
Cumulant with the following scaling function
\begin{equation}
B(L_B,T) = \mathcal{K}\left [\frac{L_B}{\xi_V(T)}\right].
\end{equation}
Note that this procedure removes the need to define any adhoc cut-off parameters to obtain the length scale. The scaling collapse looks 
quite good with the use of dynamic length scale ($\xi_V = \xi_D$) obtained in the block analysis method in the previous section. Hence we 
believe that the extracted length scale of the system $\xi_D$ for all activities will be very reliable. The results are
shown in Fig.~\ref{binder}. We also compared the Binder Cumulant of all block sizes in Fig. \ref{vanHove}c for $f_0 = 0.0$ and $f_0 = 2.5$ 
where the $\tau_\alpha$ of the both system is similar. This clearly show that the value of Binder Cumulant of van-Hove function and hence 
the associated length scale is bigger for active system than its passive counter part at similar structural relaxation timescale.

\subsection{Displacement displacement correlation function $\Gamma(r,\Delta t)$}
\begin{figure*}[!h]
\begin{center}
\vskip -0.5in
\includegraphics[width=0.7\textwidth,angle=-90]{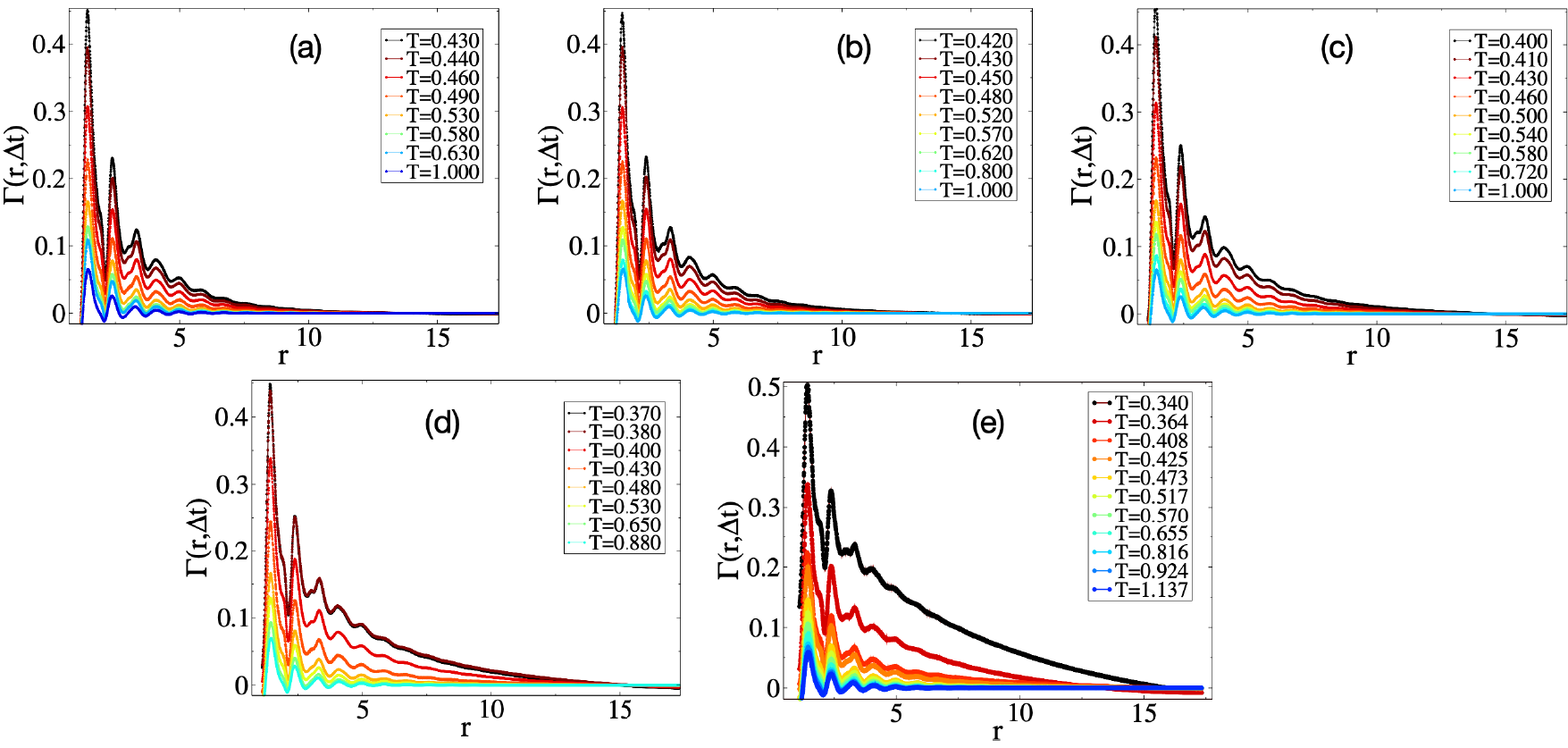}
\vskip -0.5in
\caption{excess displacement correlation $\Gamma(r,\Delta t)$ for all activities $f_0$ = 0.5 (a), 1.0 (b), 1.5 (c), 2.0 (d), 2.5 (e)}
\label{guu}
\end{center}
\end{figure*}
To understand the physical mechanism behind the enhanced dynamic heterogeneity in the active glass-forming liquids, we have measured the 
spatial correlation in the displacement field of particles at $\Delta t = \tau_\alpha$. This was already shown in previous studies to correctly capture 
the temperature dependence of the dynamic heterogeneity length scale. In this work, we have also implemented the procedure given in 
\cite{donati1999,kob1997}. The spatial correlation of the particle displacements $g_{uu}(r, \Delta t)$ is defined as
\begin{equation}
g_{uu}(r,\Delta t) = \frac{\left<\sum_{i,j=1,j\neq i}^{N}u_i(t,\Delta t)u_j(t,\Delta t)\delta(r-\mid\textbf{r}_{ij}(t)\mid)\right>}{4\pi r^2\Delta rN\rho\left<u(\Delta t)\right>^2},
\end{equation}
where $u_i(t,\Delta t) = \mid\textbf{r}_i(t+\Delta t) - \textbf{r}_i(t)\mid$ is the scalar displacement of the particle between time $t$ and $t+\Delta t$.
\begin{figure}[!h]
\begin{center}
\includegraphics[width=0.8\textwidth]{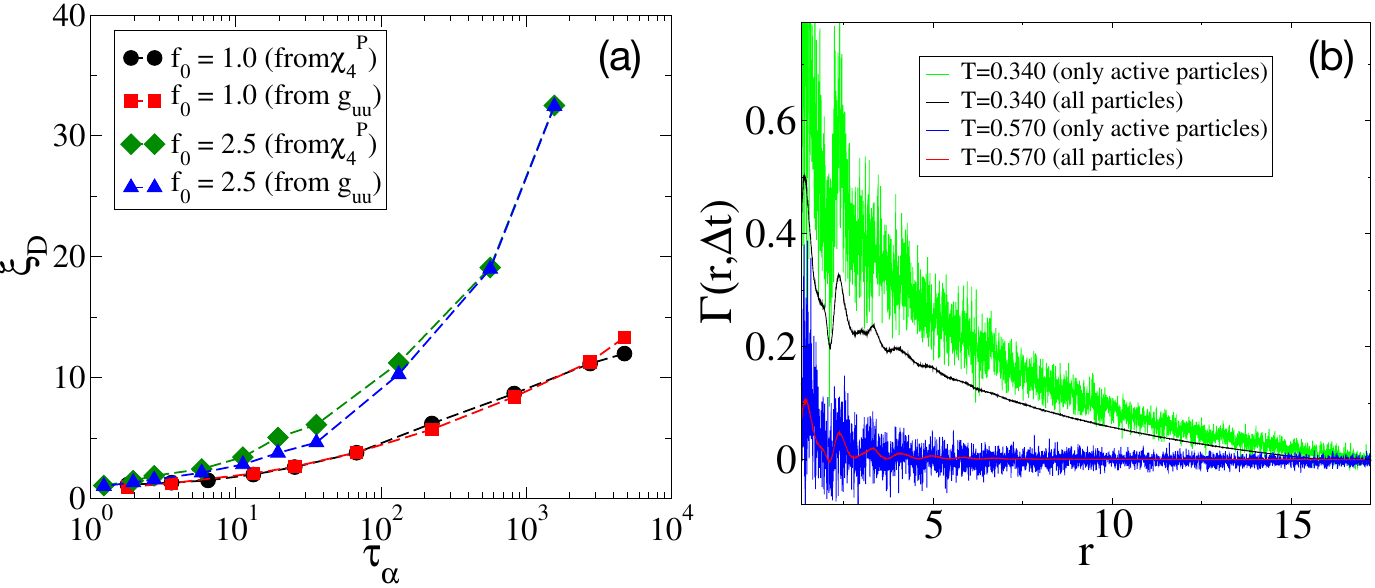}
\caption{(a) Comparison of $\xi_D$ with two different methods for $f_0 = 0.0$ and $f_0 = 2.5$. (b) Excessive displacement 
displacement correlation for full system and considering only active particles}
\label{compareXaiD}
\end{center}
\end{figure}

To extract the associated length scale we calculated the quantity called excess displacement correlation $\Gamma(r,\Delta t)$ defined as
\begin{equation}
\Gamma(r,\Delta t) = \frac{g_{uu}(r,\Delta t)}{g(r)} - 1, 
\end{equation}
where $g(r)$ is the radial pair correlation function given by
\begin{equation}
g(r) = \frac{1}{\rho N} \left<\sum_{i,j=1,j\neq i}^{N} \delta(\textbf{r} + \textbf{r}_i(0) - \textbf{r}_j(0))\right>.
\end{equation}
With this definition the excess displacement correlation function goes to zero at large distance, $\lim_{r\to\infty}\Gamma(r,\Delta t) = 0$.
We used $\Delta t = \tau_\alpha$ and calculated the excess displacement correlation $\Gamma(r,\Delta t)$ for all activities in Fig. \ref{guu}. 
The integrated area gives the associated length scale. We also compared the length scale extracted from this method with the other two methods for 
$f_0 = 0.0$ and $f_0 = 0.0$ in Fig. \ref{compareXaiD}. The extracted length scales from these different methods agree
with each other very well. This also reconfirms the robustness of the methods used to compute the dynamic heterogeneity length scale
in this study along with the reliability of the extracted length scale. Computation of length scale via different methods became necessity in 
this study as the results suggest a very dramatic rise in dynamic heterogeneity length scale with changing activity and we wanted to be 
completely sure that our observations are supported via all possible existing methods of measuring the dynamic heterogeneity length in 
the literature.

We have shown $\Gamma (r,\Delta t)$ for the highest activity $f_0 = 2.5$ at two different temperatures $T = 0.340$ ($\tau_\alpha = 1566.28$) 
and $T = 0.570$ ($\tau_\alpha = 11.195$) for $N = 50000$ in the right panel of Fig.\ref{compareXaiD}. We considered the full system as well as 
considering only 
the active particles. It is clear that the value of $\Gamma (r,\Delta t)$ is much higher if one considers only the active particles  while computing
the displacement-displacement correlation function rather than considering all the particles. This observation clearly tells that the active particles 
are probably setting up a longer range correlation in the system resulting in an enhanced dynamic heterogeneity. Understanding the origin of such 
longer range correlation with non-equilibrium active forcing will be an interesting future work and will be addressed elsewhere.

\section{Dynamic Heterogeneity and its relation with Fragility}
\begin{figure}[!h]
\begin{center}
\includegraphics[width=0.6\textwidth]{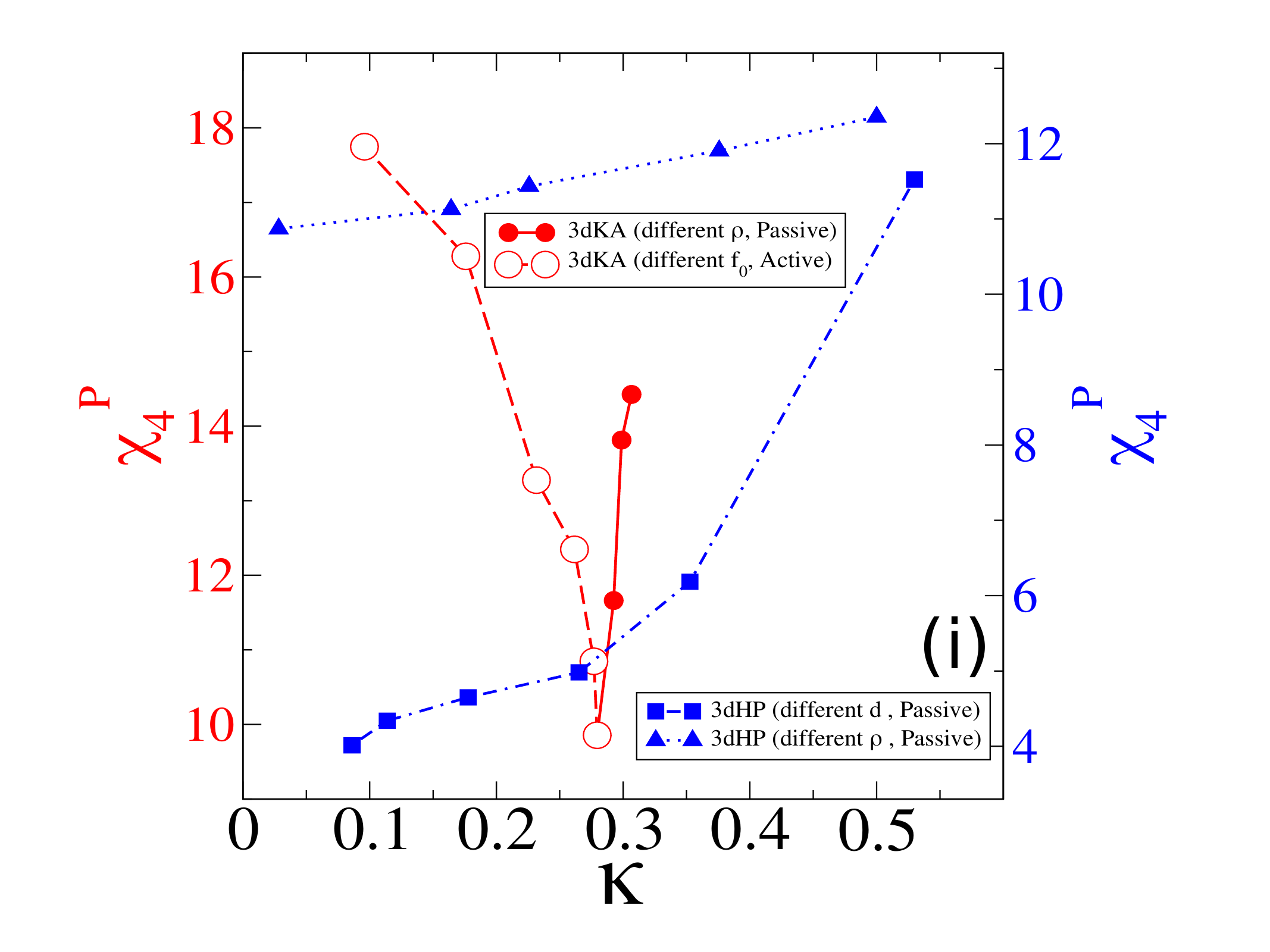}
\caption{Variation of $\chi_4^P$, as a function of fragility ($\kappa$) for equilibrium and non-equilibrium (active) glasses. Notice
that with increasing activity, fragility sharply decreases, but $\chi_4^P$ strongly increases (open circles), suggesting dramatic enhancement of heterogeneity. On the other hand, $\chi_4^P$ grows as fragility increases in equilibrium systems as denoted by red (filled circle) symbols. Blue (triangle and diamond) data sets are for 3dHP model systems taken from Refs. \cite{indrajit} and \cite{monojJPCB} respectively. It is also clear that
variation of $\chi_4^P$ shows universal behaviour with fragility in equilibrium across model systems and 
dimensions. It increases with increasing fragility irrespective of whether fragility changed due to
changes in density (Ref. \cite{indrajit}) or due to changes in the spatial dimensions (Ref. \cite{monojJPCB}).}
\label{DHFragility}
\end{center}
\end{figure}
In Figure ~\ref{DHFragility}, we show the variation of $\chi_4^P$ at a particular $\tau_\alpha$ as a function of fragility, $\kappa$ to highlight the striking effect of the non-equilibrium active driving force leading to qualitative differences between the physics of equilibrium and active glasses. Here $\kappa$ is called the kinetic fragility and is defined from the well known Vogel-Fulcher-Tamann formula as 
\begin{equation}
\kappa = \frac{1}{\ln(\frac{\tau_\alpha}{\tau_0}) [\frac{T}{T_{VFT}} - 1]},
\end{equation}
where $\tau_0$ is the viscosity at infinite high temperature and $T_{VFT}$ is the apparent divergence temperature for relaxation time. As evident, $\chi_4^P$ monotonically decreases as $\kappa$ increases in active glasses, whereas it shows the opposite trend in equilibrium glasses. Dynamic heterogeneity in equilibrium seems to have a universal relationship with fragility: it increases with increasing fragility in equilibrium glasses irrespective of how the fragility changes in a wide variety of systems. For example, the 3dKA model shows an increase in fragility with increasing density (red circle); the same is true for the HP model (blue triangle) in three dimensions (data taken from Ref.\cite{indrajit}). Moreover, the HP model also shows a significant change in fragility with changing spatial dimensions (blue diamond) (data taken from Ref.\cite{monojJPCB}) at a specific relative density from the jamming density. Nevertheless, in all these scenarios, $\chi_4^P$ increases universally with increasing fragility in sharp contrast with results obtained in this studied non-equilibrium glasses where $\chi_4^P$ decreases as fragility increases. It remains unclear whether this latter behavior is universal for active glasses. However, it is clear that the marked deviation from equilibrium behavior is solely due to the non-equilibrium driving forces in this active glass model.

\subsection*{Dependence of $\xi_D$ with $T_g/T$}	
\begin{figure*}[!h]
\begin{center}
\includegraphics[width=0.6\textwidth]{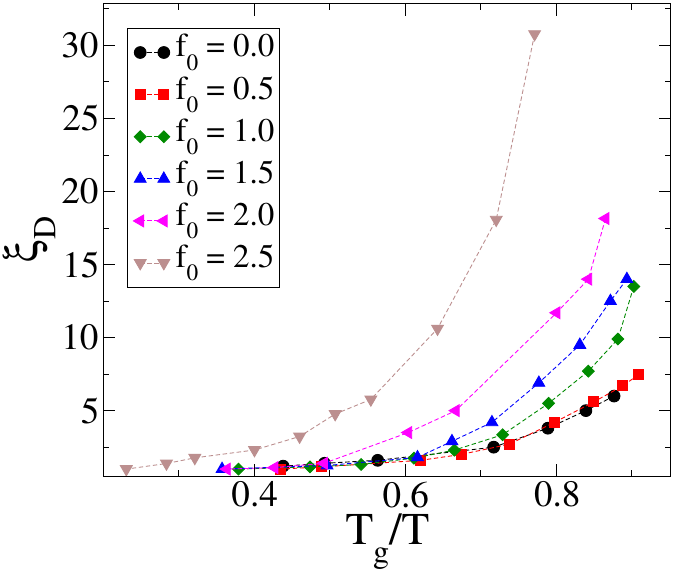}
\caption{Dynamic heterogeneity length scale $\xi_D$ as a function of $T_g/T$ to demonstrate the dramatic growth of length scale within a comparable range of $T_g/T$, where $T_g$ is the calorimetric glass transition temperature, defined as $\tau_\alpha(T_g)=10^6$.}
\label{xaiDvsTgbyT}
\end{center}
\end{figure*}

We have also presented the dynamic heterogeneity length scale, $\xi_D$, as a function of $T_g/T$ in Fig. \ref{xaiDvsTgbyT}, where $T_g$ is the calorimetric glass transition temperature, defined as $\tau_\alpha(T=T_g)=10^6$. It clearly shows the dramatic increase of $\xi_D$ with increasing $f_0$. 



\begin{figure}[!h]
	\includegraphics[width=\textwidth]{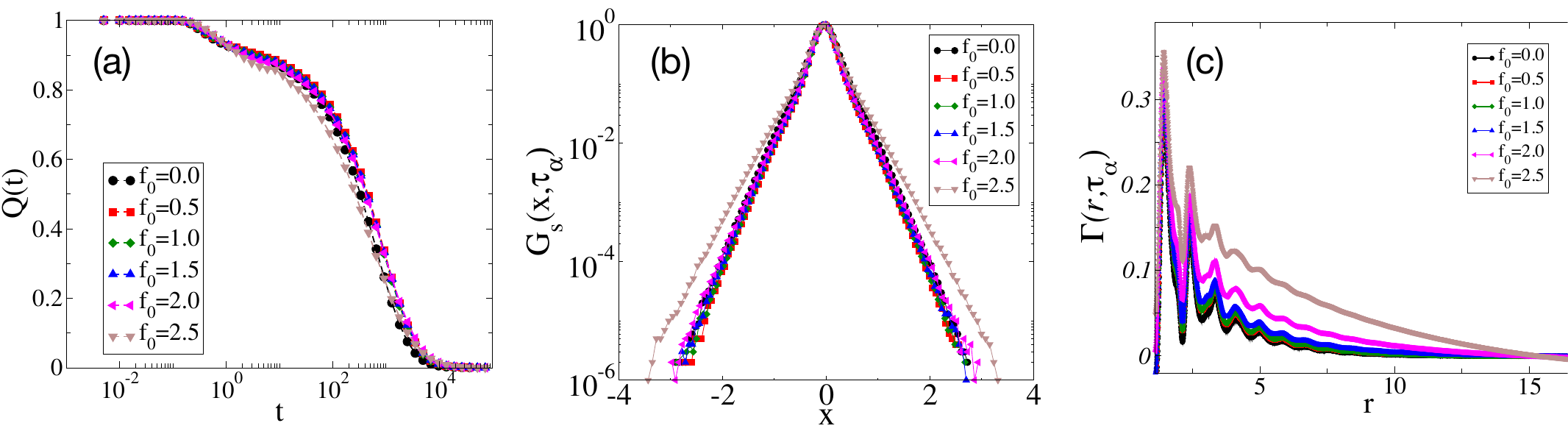}
	\caption{(a) $T$ and $f_0$ are chosen such that $Q(t)$ almost overlaps, showing similar $\tau_{\alpha}$. (b) The van Hove function at those set of parameters as in (a) shows higher non-Gaussianity as $f_0$ increases. (c) Increasing spatial correlation, as measured by the excess displacement-displacement correlation function, $\Gamma(r,\tau_\alpha)$, with increasing activity. These correlation functions are
again computed at the same $\tau_\alpha$ to highlight the growth of spatial heterogeneity with increasing activity}
	\label{SMvanHove}
\end{figure}

\section{Higher non-Gaussian behaviour at larger activity}
We have shown in the main text that the dynamical heterogeneity grows as activity increases. Similar effect can also be found in other dynamical quantities as well.
Figure~\ref{SMvanHove}(b) shows the increasing non-Gaussian behaviour in the van Hove function, $G_s(x,\tau_\alpha)$, with increasing activity while $\tau_\alpha$ remains constant, as the plots of $Q(t)$ confirm (Fig. \ref{SMvanHove}(a)). This shows growing DH with increasing activity even when $\tau_\alpha$ is same.
Figure~\ref{SMvanHove}(c) once more demonstrates that spatial correlation, computed by the excess part of the displacement-displacement correlation function, $\Gamma(r,\tau_\alpha)$, increases markedly with increasing activity, confirming the strong decoupling of DH and structural relaxation dynamics in active glasses.

\section{Correlation between $\tau_\alpha$ and $\tau_{peak}$}
\begin{figure}[!h]
\begin{center}
\includegraphics[scale=0.5]{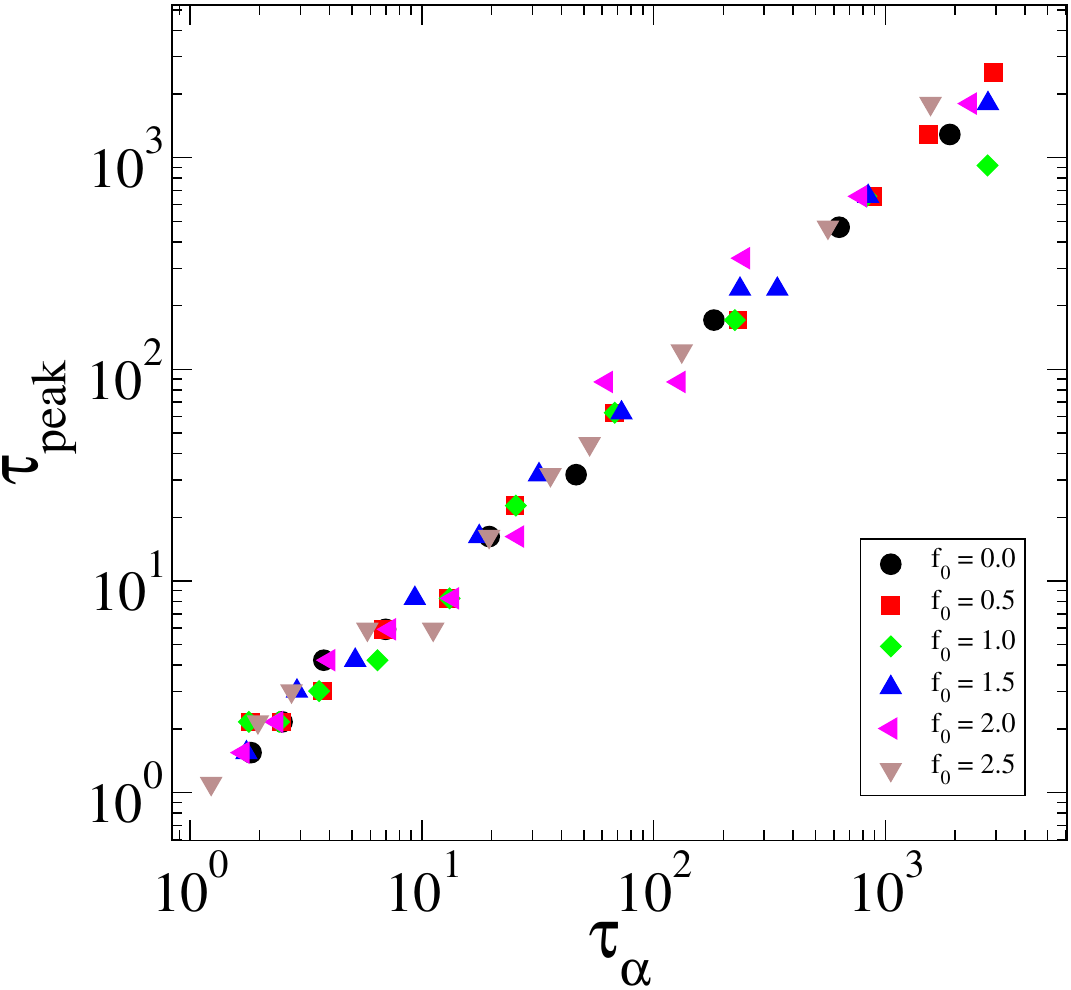}
\caption{The cross plot of $\tau_\alpha$ and $\tau_{peak}$ for all the studied activity and temperature range. The nice data collapse 
confirms that the time at which the peak appears in $\chi_4(t)$ is proportional to the $\alpha$-relaxation time of the system similar 
to the passive case.}
\label{tauAlphavsTauPeak}
\end{center}
\end{figure}

The time, $\tau_{peak}$, at which $\chi_4(t)$ attains its maximum gives a measure of relaxation time. Figure \ref{tauAlphavsTauPeak} shows $\tau_{peak}$ against  $\tau_\alpha$, relaxation time obtained from $Q(t)$, of the system for all studied $T$ and $f_0$. Near collapse of the data confirms that $\tau_{peak}$ is proportional to $\tau_\alpha$ even in the presence of active forces. This observation, along with the analysis of $\tau_\alpha$ presented in the main text, suggest that the relaxation dynamics, characterized via either $\chi_4(t)$ or $Q(t)$, can be understood by an effective-equilibrium-like description at an appropriate effective temperature.

\section{Cooperatively rearranging region (CRR)}
\begin{figure}[!h]
\begin{center}
\includegraphics[scale=0.5]{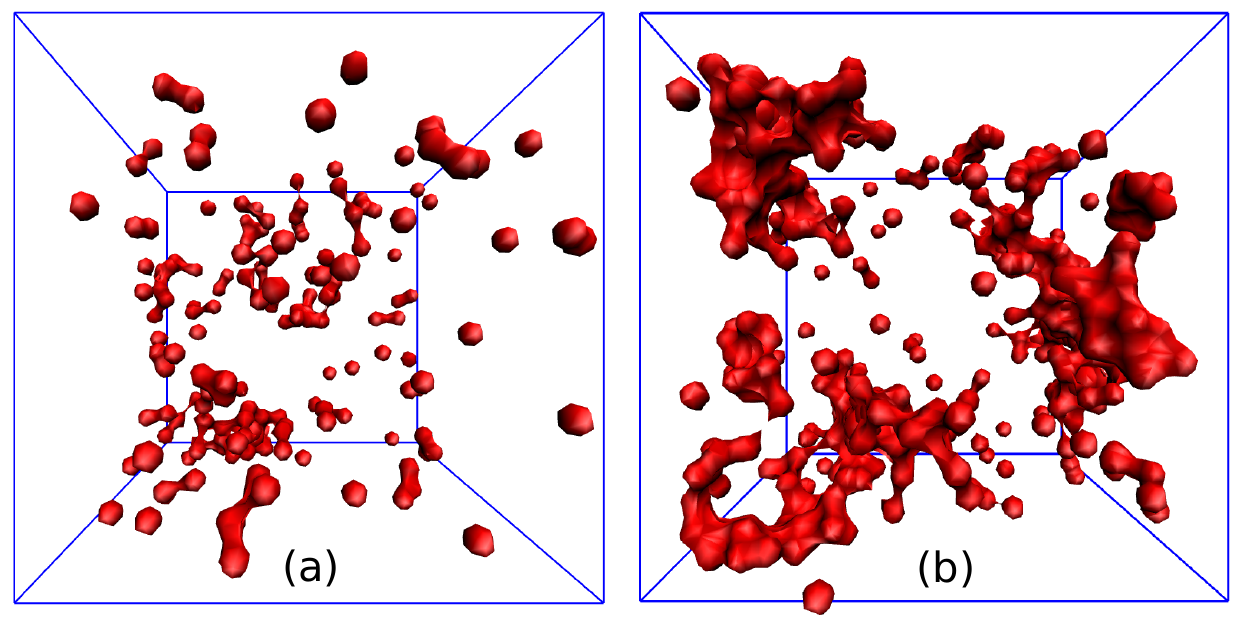}
\caption{Clustering of faster particles at same $\tau_\alpha=10^3$ for (a) $f_0 = 0.0$ and (b) $f_0 = 2.5$.}
\label{CRR}
\end{center}
\end{figure}
The dynamic heterogeneity scenario of glassy dynamics refers to the coexistence of dynamically slow and fast relaxing regions in the system. A region consisting of neighboring particles with comparable relaxation time relaxes collectively and known as cooperatively rearranging region (CRR). Thus,  CRR provides a qualitative measure of dynamic heterogeneity in the system. To observe this region in three dimension, both for the passive and the active systems, we calculate the net displacement of each particles over a time-scale $\tau_\alpha$ ($\sim10^3$). We then consider only the faster particles whose displacements are greater than our chosen cut off value which is 9\% of the whole simulation box. 
We plot the positions of these particles in the VMD software using the function ``surf'': each particle is represented by a small sphere of certain radius and if the distance between two particles are less than their diameter, they are collectively represented by a surface.
Figure \ref{CRR}(a) shows a typical plot of the faster paricles for equilibrium glass while Fig. \ref{CRR}(b) shows the same for $f_0 = 2.5$; the parameters of the two systems are chosen such their $\tau_\alpha$ are same.
It is clear from the plot that unlike the equilibrium system, the CRR in case of active system is system spanning. This visualization gives a qualitative idea of enhanced dynamic heterogeneity in active glass compared to its equilibrium counterpart. \\

\begin{figure}[!h]
	\begin{center}
		\includegraphics[scale=0.75]{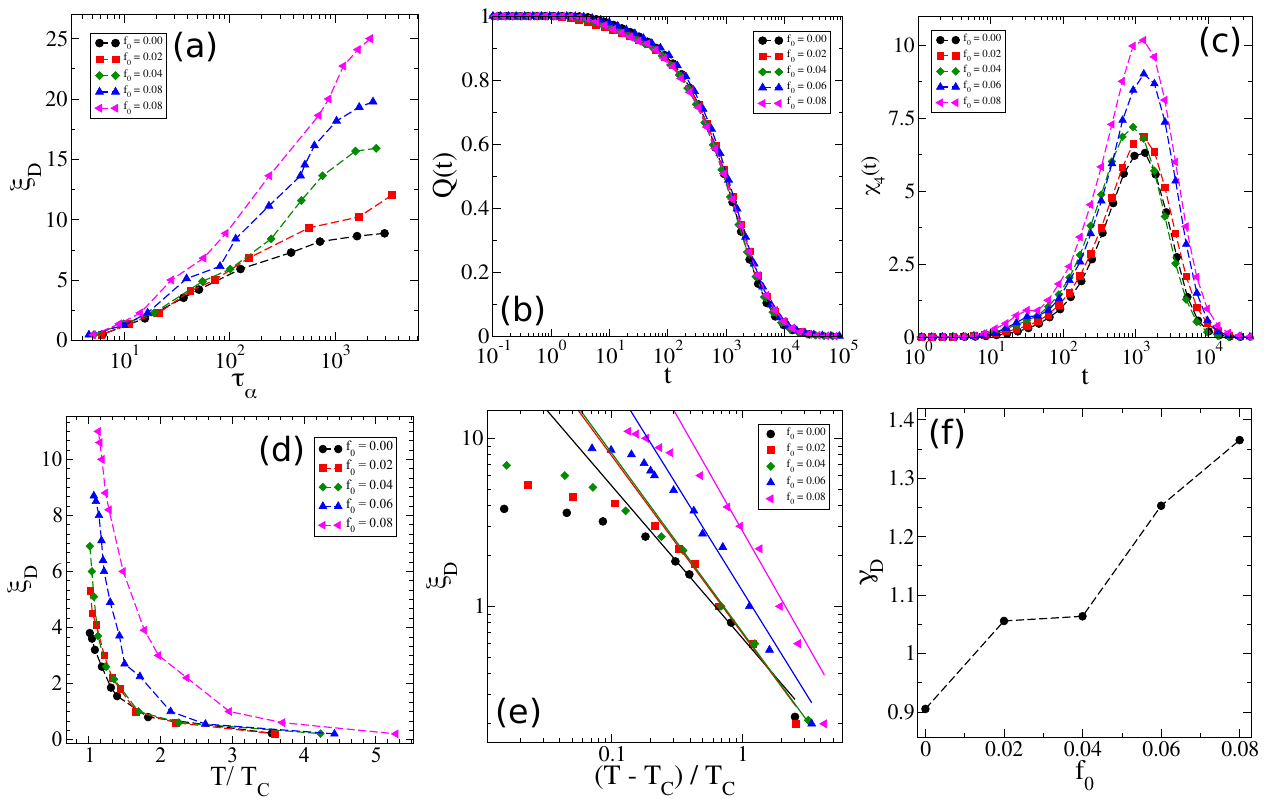}
		\caption{Simulation results for the 3dHP model: (a) $\xi_D$ as a function of $\tau_\alpha$ for different $f_0$. (b) Overlap correlation function, $Q(t)$, for different activities where the systems have similar $\tau_\alpha$. (c) $\chi_4(t)$  for the same systems as in (b). (d) $\xi_D$ as a function of $T/T_C$. (e) $\xi_D$ as a function of $(T-T_C)/T_C$, lines are fits to the form $\xi_D\sim [(T-T_C)/T_C]^{-\gamma_D}$. (f) The exponent $\gamma_D$ increases almost linearly with $f_0$.}
		\label{3dHP}
	\end{center}
\end{figure}

\section{Results of 3dHP}
In this section, we present some of the main results obtained from 3dHP model system which confirm that the results presented in the main manuscript are generic and applicable to a wide class of model systems. Note that this model is very different from the 3dKA model
which is a good model for molecular glass-forming liquids, whereas 3dHP is a paradigmatic model for soft sphere systems relevant for colloidal systems and has been widely studied in the context of jamming physics. The dynamic heterogeneity length scale $\xi_D$ in this model is also computed in the same way as discussed before for 3dKA model. For a better comparison, we have used the same $\tau_p=1.0$ in the simulations of the 3dHP model. The results are very similar as 3dKA model, presented in the main text. In 
Fig. \ref{3dHP} (a) we have shown the dynamic heterogeneity length scale $\xi_D$ as a function of $\tau_\alpha$ where the length scale 
at a particular $\tau_\alpha$ is increasing with increasing $f_0$. In Fig. \ref{3dHP} (b) and (c) we have shown the overlap correlation function, $Q(t)$, 
and the four point susceptibility, $\chi_4(t)$. The parameters in these systems are chosen such that they all have similar $\tau_\alpha$, as confirmed from the plot fo $Q(t)$ [Fig. \ref{3dHP}(b)]. It is clear that the peat height, $\chi_4^P$, of $\chi_4(t)$ is 
monotonically increasing with increasing activity which has similar trend as 3dKA model. In Fig. \ref{3dHP} (d) and (e) we have shown 
$\xi_D$ as function of $T/T_C$ and $(T-T_C)/T_C$ respectively. There is a power law behaviour observed between $\xi_D$ and the 
rescaled temperature as $\xi_D(T) \propto \left|\frac{T-T_C}{T_C}\right|^{-\gamma_D}$ with an exponent $\gamma_D$. The variation of this
exponent $\gamma_D$ is shown in the panel (f).  $\gamma_D$ seems to linearly increase with $f_0$, similar to the result in simulations of 3dKA model (Fig. \ref{xaiDvsTbyTk-1}).

\section{Active Inhomogeneous Mode-Coupling Theory (active-IMCT)}
As noted in the main text, we calculate the four-point correlation function, $\chi_4(t)$, via linear-response theory by calculating the corresponding susceptibility under a weak external field, $\u(\br,t)$, that couples to the density, $\rho(\br,t)$.
The hydrodynamic equations of motion for $\rho(\br,t)$ and the momentum density, $\rho(\br,t)\mathbf{v}(\br,t)$, for an active fluid are
\begin{align}
\f{\p\rho(\br,t)}{\p t}&=-\nabla\cdot (\rho(\br,t)\mathbf{v}(\br,t)) \label{contrho}\\
\f{\p (\rho(\br,t)\mathbf{v}(\br,t))}{\p t}+\nabla\cdot(\rho(\br,t)\mathbf{v}(\br,t)\mathbf{v}(\br,t)) &=\eta \nabla^2\mathbf{v}(\br,t)+\left(\zeta+\f{\eta}{3}\right)\nabla\nabla\cdot\mathbf{v}(\br,t)-\rho(\br,t)\nabla\f{\delta \fu}{\delta\rho(\br,t)}+\mathbf{f}_T(\br,t)+\mathbf{f}_A(\br,t), \label{contmo}
\end{align}
where $\eta$ and $\zeta$ are bulk and shear viscosities, respectively, $\mathbf{f}_T(\br,t)$, the thermal noise with zero mean and 
\begin{equation}
\langle \mathbf{f}_T(\mathbf{0},0)\mathbf{f}_T(\br,t)\rangle=-2k_BT\left[ \eta \mathbf{I}\nabla^2+\left(\zeta+\f{\eta}{3}\right)\nabla\nabla \right]\delta(\br)\delta(t)
\end{equation}
with $k_BT$ being the Boltzmann constant times temperature, $\mathbf{I}$, the unit tensor in spatial dimension $d$. Activity in our system enters via the active noise, $\mathbf{f}_A(\br,t)$, with zero mean and 
\begin{equation}
\langle \mathbf{f}_{A}(\br,t)\mathbf{f}_A(\mathbf{0},0)\rangle=2\Delta(\br,t)\mathbf{I},
\end{equation}
where $\Delta(\br,t)$ specifies the type of activity. $\fu$ is a free energy functional in the presence of $\u$
\begin{align}
\beta\fu=\int_{\br}\left[\rho(\br,t)\ln\left(\f{\rho(\br,t)}{\rho_0}\right)-\delta\rho(\br,t)\right]-\f{1}{2}\int_{\br,\br'} c(\br-\br')\delta\rho(\br,t)\delta\rho(\br',t)+\beta\int_\br \u(\br)\delta\rho(\br,t),
\end{align}
where $\beta=1/k_BT$, the static density, $m(\br)$, becomes inhomogeneous in the presence of $\u(\br)$ and we have $\rho(\br,t)=m(\br)+\dr$, where $\dr$ denotes the fluctuating part of density. We have used the notation $\int_\br\equiv \int \md\br$. The direct correlation function, $c(\br-\br')$, can be anisotropic in the presence of $\u$. Minimization of $\fu$ gives $m(\br)$,
\begin{equation}
\beta\f{\delta\fu}{\dr}\bigg|_{\rho(\br,t)=m(\br)}=0.
\end{equation}

Now, the static inhomogeneous density has two contributions, $\rho_0$, the homogeneous background density in the absence of $\u$, and $\delta m(\br)$, the static density due to the presence of $\u$ alone. Thus, $m(\br)=\rho+\delta m(\br)$, where $\delta m(\br)$ is small as $\u$ is weak. Both the density fluctuation, $\dr$, and the velocity field are small in the glassy regime. Therefore, we linearize Eqs. (\ref{contrho}) and (\ref{contmo}), by neglective $\dr\mathbf{v}(\br,t)$ and higher-order terms. We disregard both the acceleration and inertia terms in Eq. (\ref{contmo}), take a divergence and replace $\nabla\cdot\mathbf{v}(\br,t)$ using Eq. (\ref{contrho}). This necessitates the evaluation of $\nabla\cdot\left(\rho(\br,t)\nabla(\delta \beta\fu/\dr)\right)$. The force density, within linear order in $\delta m(\br)$ is obtained as
\begin{align}
\rho(\br,t)\nabla\f{\delta \fu}{\dr}&=\nabla\int_\br[\delta(\br-\br')-\rho_0c(r-r')]\delta\rho(\br',t)-\nabla\int_{\br'}\delta m(\br)c(r-r')\delta\rho(\br',t) \nonumber\\
&-\dr\nabla\int_{\br'}c(r-r')\delta\rho(\br',t)-\f{\nabla\delta m(\br)}{\rho_0}[\delta(\br-\br')-\rho_0c(r-r')]\delta\rho(\br',t),
\end{align}
where we have written $c(r-r')$, the equilibrium direct correlation function, to emphasize that the anisotropy is explicitly included via $\delta m(\br)$. Since we are going to write the equation of motion for the density fluctuation in Fourier space with wavevector $\bk$, we evaluate the divergence of the force density in $\bk$-space as
\begin{align}\label{divforce}
\left[\nabla\cdot\left(\rho(\br,t)\nabla\f{\delta\beta\fu}{\dr}\right)\right]_\bk&=-k^2k_BT (1-\rho_0c_k)\drk_\bk(t)+k^2k_BT\int_\bq\delta m_{\bk-\bq}c_q\drk_\bq(t) \nonumber\\
&+\f{k_BT}{\rho_0}\int_\bq\frac{\bk\cdot(\bk-\bq)}{S_q}\delta m_{\bk-\bq}\drk_\bq(t)+\f{k_BT}{2}\int_\bq \bk\cdot[\bq c_q+(\bk-\bq)c_{k-q}]\drk_\bq(t)\drk_{\bk-\bq}(t),
\end{align}
where $c_k$ and $S_k$ are the direct correlation function and static structure factor, respectively, in Fourier space. Within MCT, the glass transition is a critical phenomenon with a diverging length scale. Therefore, our interest is the $\bk\to0$ limit and $\u$ is chosen such that $\delta m_\bk$ is sharply peaked at around $\bk\to0$. Then, Eq. (\ref{divforce}) for this particular choice of $\u$ becomes
\begin{align}
\left[\nabla\cdot\left(\rho(\br,t)\nabla\f{\delta\beta\fu}{\dr}\right)\right]_\bk=-\f{k^2k_BT}{S_k}\drk_\bk(t)+k^2k_BT\delta m_{\bf 0}c_k\drk_\bk(t)+\f{k_BT}{2}\int_\bq \bk\cdot[\bq c_q+(\bk-\bq)c_{k-q}]\drk_\bq(t)\drk_{\bk-\bq}(t).
\end{align}
Using the above expression, we obtain the equation of motion for density fluctuation in Fourier space as
\begin{align}\label{lankspace}
D_Lk^2\f{\p\drk_\bk(t)}{\p t}&+\f{k_BTk^2}{S_k}\drk_\bk(t)-k^2k_BT\delta m_{\bf 0}c_k\drk_\bk(t) \nonumber\\
&=\f{k_BT}{2}\int_\bq \bk\cdot[\bq c_q+(\bk-\bq)c_{k-q}]\drk_\bq(t)\drk_{\bk-\bq}(t)+ik\hat{f}_T^L(\bk,t)+ik\hat{f}_A^L(\bk,t),
\end{align}
where $\hat{f}_T^L(\bk,t)$ and $\hat{f}_A^L(\bk,t)$ are the longitudinal parts of the thermal and active noises in Fourier space and $D_L=(\zeta+4\eta/3)/\rho_0$. Equation (\ref{lankspace}) can now be treated within the field-theoretic technique \cite{reichman2005,saroj2016} to obtain the equations of inhomogeneous mode-coupling theory for an active fluid. However, the numerical solution of the wavevector-dependent theory, even for the two-point correlation functions, is impractical due to a high computation time requirement \cite{activemct}. In such cases, simplified integral equations ignoring the wavevector dependence and keeping track of the time-dependence alone have been invaluable in extracting meaningful insights from MCT within a manageable calculation. Such a version of the theory is traditionally known as schematic MCT. Thus, we take a schematic limit of the Langevin equation, Eq. (\ref{lankspace}), throwing away the wavevector dependence, and obtain
\begin{equation}\label{langevineq}
\f{\p\phi(t)}{\p t}+\mu(t)\phi(t)=\f{g}{2}\phi(t)^2+\be \phi(t)+f_T(t)+f_A(t),
\end{equation}
where $\phi(t)$ is the density fluctuation, $f_T(t)$ is the thermal noise with $\langle f_T(t)f_T(t')\rangle=2T\delta(t-t')$ and $f_A(t)$ is the active noise with $\langle f_A(t)f_A(t')\rangle=2\Delta(t-t')$. The nature of the active noise depends on the type of system. Two models of activity were considered in Refs. \cite{activemct,activerfot} and both of them have been extensively used in recent simulation works \cite{mandal2016,flenner2016}. In the current work, we consider $\Delta(t)=f_0^2\exp[-t/\tau_p]$, where $f_0$ is the self-propulsion force and $\tau_p$ is the persistence time. Note that the difference in the two types of noise statistics is not relevant here since we probe the effect of activity via $f_0$ keeping $\tau_p$ constant; both the noise statistics are identical in this scenario. $\be$ gives the effect of the external field, $\mu(t)$ is the frequency term, and $g$ denotes the interaction strength. We now use Eq. (\ref{langevineq}) as a starting point of a field-theoretic treatment to obtain the mode-coupling theory. 

The correlation and response functions are defined as $Q(t,t')=\langle \phi(t)\phi(t')\rangle$ and $R(t,t')=\langle\p\phi(t)/\p f_T(t')\rangle$ respectively; the corresponding functions in the presence of the external field are defined as $\tilde{Q}(t,t')$ and $\tilde{R}(t,t')$. Going through the standard field-theoretic approach \cite{activemct,saroj2016,reichman2005} and writing $g^2=4\lambda$, we obtain
\begin{align}
\f{\p \tilde{Q}(t,t')}{\p t} &=-\mu(t)\tc(t,t')+\int_0^{t'}\md s\mathcal{D}(t,s)\tilde{R}(t',s)+\int_0^t \md s \mathcal{S}(t,s)\tc(s,t')+2T\tilde{R}(t',t)+\be \tc(t,t')\\
\f{\p \tilde{R}(t,t')}{\p t} &=\delta(t-t')-\mu(t)\tilde{R}(t,t')+\int_{t'}^t\md s \mathcal{S}(t,s)\tilde{R}(s,t')+\be \tilde{R}(t,t'),
\end{align}
where $\mathcal{D}(t,s)=2\lambda \tc^2(t,s)+\Delta(t-s)$ and $\mathcal{S}(t,s)=4\lambda \tc(t,s)\tilde{R}(t,s)$.
Without losing generality, we now consider $t>t'$, and thus, the response function $\tilde{R}(t',t)$ becomes zero due to the boundary condition obeying causality. For the numerical solution, it is advantageous to define an integrated response function, $\tilde{F}(t,t')$, as
\begin{equation}
\tilde{F}(t,t')=-\int_{t'}^t\tilde{R}(t,s)\md s.
\end{equation}

We are interested in the steady state dynamics of the active system. Therefore, we take the limit $t\to\infty$ and $t'\to\infty$ such that $(t-t')$ remains constant, then $\tc(t,t')$ and $\tilde{F}(t,t')$ become functions of the time difference alone, i.e., 
\begin{equation}
\tc(t,t')\equiv \tc(t-t'); \,\,\,  \tilde{F}(t,t')\equiv \tilde{F}(t-t').
\end{equation}
We redefine $(t-t')\to t$ and, following Ref. \cite{activemct}, write down the MCT equations for the active system in the presence of the externally applied field as
\begin{subequations}
	\label{twopt_field}
	\begin{align}
	\f{\p \tc(t)}{\p t} &= \Pi(t)-(T-p)\tc(t)-\int_0^tm(t-s)\f{\p\tc(s)}{\p s}\md s+\be\tc(t)\\
	\f{\p\tilde{F}(t)}{\p t} &= -1 -(T-P)\tilde{F}(t)-\int_0^t m(t-s)\f{\p \tilde{F}(s)}{\p s}\md s+\be \tilde{F}(t),
	\end{align}
\end{subequations}
with,
\begin{align}
&\Pi(t) =-\int_t^\infty \Delta(s) \f{\p \tilde{F}(s-t)}{\p s}\md s,\\
&p = \int_0^\infty \Delta(s)\f{\p \tilde{F}(s)}{\p s}\md s \\
\text{and} \hspace{2cm} &m(t-s) =2\lambda \f{\tc^2(t-s)}{T_{eff}(t-s)} \label{memory_field}
\end{align}
where $T_{eff}(t)$ is defined through a generalized fluctuation-dissipation relation (FDR) for nonequilibrium systems as
\begin{equation}
\f{\p \tc(t)}{\p t}=T_{eff}(t)\f{\p \tilde{F}(t)}{\p t}.
\end{equation}

Following Ref. \cite{IMCT}, we now define the four-point correlation functions, corresponding to $C(t)$ and $F(t)$, as
\begin{equation}
\chi_Q(t)=\f{\p \tc(t)}{\p \be}\bigg|_{\be\to0} \,\,\, \text{and}\,\,\, \chi_F(t)=\f{\p \tilde{F}(t)}{\p \be}\bigg|_{\be\to0}.
\end{equation}
Thus, the equations of motions for $\chi_Q(t)$ and $\chi_F(t)$ are readily obtained from Eqs. (\ref{twopt_field}) as
\begin{subequations}
	\label{chieq}
	\begin{align}
	\f{\p\chi_Q(t)}{\p t} &=\nu(t)+(1+\zeta)Q(t)-(T-p)\chi_Q(t)-\int_0^tm(t-s)\f{\p \chi_Q(t)}{\p s}\md s-\int_0^t \Sigma(t-s)\f{\p Q(s)}{\p s}\md s \\
	\f{\p\chi_F(t)}{\p t} &=(1+\zeta)F(t)-(T-p)\chi_F(t)-\int_0^tm(t-s)\f{\p \chi_F(t)}{\p s}\md s-\int_0^t \Sigma(t-s)\f{\p F(s)}{\p s}\md s,
	\end{align}
\end{subequations}
where
\begin{align}
\nu(t)&=-\int_t^\infty \Delta(s) \f{\p\chi_F(s-t)}{\p s}\md s \\
\zeta&=\int_0^\infty \Delta(s) \f{\p\chi_F(s)}{\p s}\md s\\
\text{and} \,\,\, \Sigma(t)&=2\lambda\f{Q(t)\chi_Q(t)}{T_{eff}(t)}+2\lambda\f{Q(t)^2}{T_{eff}(t)^2}\kappa f_0^2, \label{sigmaterm}
\end{align}
along with the equations for the two-point correlation functions
\begin{subequations}
	\begin{align}
	\f{\p Q(t)}{\p t} &= \Pi(t)-(T-p)Q(t)-\int_0^tm(t-s)\f{\p Q(s)}{\p s}\md s\\
	\f{\p{F}(t)}{\p t} &= -1 -(T-P){F}(t)-\int_0^t m(t-s)\f{\p F(s)}{\p s}\md s,
	\end{align}
\end{subequations}
with,
\begin{align}
&\Pi(t) =-\int_t^\infty \Delta(s) \f{\p {F}(s-t)}{\p s}\md s,\\
&p = \int_0^\infty \Delta(s)\f{\p {F}(s)}{\p s}\md s \\
&m(t-s) =2\lambda \f{Q^2(t-s)}{T_{eff}(t-s)} \\
\text{and} \hspace{2cm} &\f{\p Q(t)}{\p t}=T_{eff}(t)\f{\p {F}(t)}{\p t}. \label{teff_final}
\end{align}
$\kappa$ is a constant since $\tau_p$ in our simulation is fixed. $\kappa$ essentially depends on the system-dependent parameters of how activity is related to the long-time value of $T_{eff}(\infty)$. As shown in Ref. \cite{activemct}, this dependence can be calculated within the approximation of a single active particle inside a potential. As shown in Ref. \cite{IMCT}, $\chi_Q(t)\equiv \chi_4(t)$. Note the nontrivial contribution from the activity is encoded in the second term in Eq. (\ref{sigmaterm}), this is a consequence of having $T_{eff}$, instead of $T$, in the denominator of Eq. (\ref{memory_field}). 

Equations (\ref{chieq})-(\ref{teff_final}) give the IMCT for an active glass.
An analytical solution of the IMCT equations is not possible, and one must solve them numerically. Even then, the numerical solution of the wavevector-dependent theory is impractical as it requires a long computational time. The control parameters, such as $T$ and density, are given by the parameter $\lambda$ in the schematic MCT. Numerical solutions of the theory at different values of $\lambda$ and $f_0$ are shown in the main article.

%

\bibliography{activechi4_SIref.bib}